\documentclass[12pt,preprint]{aastex}
\usepackage{natbib}
\usepackage{epsf}

\def\percc{\mbox{$\;{\rm cm}^{-3}$}}
\def\permpc{\mbox{$\;{\rm Mpc}^{-1}$}}
\def\kms{\mbox{$\;$km s$^{-1}$}}     
\def\kmsec{\mbox{$\;$km s$^{-1}$}}     
\def\cms2{\mbox{$\;$cm s$^{-2}$}}    
\def\ergs{\mbox{$\;$ergs}}
\def\ergss{\mbox{$\;$ergs s$^{-1}$}}

\def\mb{\mbox{$\;{\rm mbarn}$}}
\def\mug{\mbox{${\rm {\mu}G}$}}
\def\msun{\mbox{${{\rm M}_\odot}$}}

\def\arcsec{\rlap{$^{\prime\prime}$}\hbox to 2pt{}}
\def\arcmin{\rlap{$^{\prime}$}\hbox to 2pt{}}
\def\H0{\mbox{${\rm H_{0}}$}}
\def\mpc{\mbox{$\;{\rm Mpc}$}}
\def\kpc{\mbox{$\;{\rm kpc}$}}
\def\mev{\mbox{$\;{\rm MeV}$}}
\def\mevs{\mbox{$\;{\rm MeV}\; {\rm s}^{-1}$}}
\def\mevcc{\mbox{$\;{\rm MeV\; {\rm cm}^{-3}}$}}
\def\flux{\mbox{$\;{\rm ergs\; {\rm cm}^{-2}\; {\rm s}^{-1}}$}}
\def\kev{\mbox{$\;{\rm keV}$}}
\def\gev{\mbox{$\;{\rm GeV}$}}
\def\ev{\mbox{$\;{\rm eV}$}}

\def\gyr{\mbox{$\;{\rm Gyr}$}}
\def\yrs{\mbox{$\;{\rm yrs}$}}

\def\etal{{\it {et al.}}}
\def\gtorder{\mathrel{\raise.3ex\hbox{$>$}\mkern-14mu
             \lower0.8ex\hbox{$\sim$}}}
\def\ltorder{\mathrel{\raise.3ex\hbox{$<$}\mkern-14mu
             \lower0.8ex\hbox{$\sim$}}}

\begin{document}
\bibliographystyle{plain}
\hyphenation{brems-strah-lung}

\title{Nonthermal Particles and Radiation Produced by Cluster Merger Shocks}
\author{Robert C. Berrington} 
\affil{Naval Research Lab, Code 7653, 4555 Overlook SW, Washington, DC,
  20375-5352}
\email{rberring@gamma.nrl.navy.mil}
\author{Charles D.\ Dermer}
\affil{Naval Research Lab, Code 7653, 4555 Overlook SW, Washington, DC,
  20375-5352} 
\email{dermer@gamma.nrl.navy.mil}


\begin{abstract}
We have developed a numerical model for the temporal evolution of particle and
photon spectra resulting from nonthermal processes at the shock fronts formed
in merging clusters of galaxies.  Fermi acceleration is approximated by
injecting power-law distributions of particles during a merger event, subject
to constraints on maximum particle energies. We consider synchrotron,
bremsstrahlung, Compton, and Coulomb processes for the electrons, nuclear,
photomeson, and Coulomb processes for the protons, and knock-on electron
production during the merging process.  The broadband radio through
$\gamma$-ray emission radiated by nonthermal protons and primary and secondary
electrons is calculated both during and after the merger event.  Using ROSAT
observations to establish typical parameters for the matter density profile of
clusters of galaxies, we find that typical merger shocks are weak and
accelerate particles with relatively soft spectra. We consider the prospects
for detecting nonthermal radio and $\gamma$-ray emission from clusters of
galaxies and implications for the origin of ultra-high energy cosmic rays and
the diffuse $\gamma$-ray background. Our results suggest that only a few of
the isotropically-distributed unidentified EGRET sources are due to shocks
formed in cluster mergers, and that only a minor contribution to the diffuse
extragalactic $\gamma$-ray background can originate from cluster merger
shocks. Cluster merger shocks can accelerate protons to $\lesssim 10^{19}$ eV
for the standard parameters considered here. We predict that {\it GLAST} will
detect several cluster mergers and, depending on the mean magnetic fields in
the intracluster medium, the Low Frequency Array could detect anywhere from
several to several hundred.
\end{abstract}

\section{Introduction}
\label{sec:introduction}

According to the hierarchical merging scenario, cold dark matter halos evolve
to form larger structures by merging with adjacent dark matter halos.  Within
dark matter halos, baryonic matter condenses to form clusters of galaxies.  A
cluster merger event results from the interaction of galaxy clusters during
the merger of cold dark matter halos.  The gravitational potential energy
available in a cluster merger event involving halos with masses $\sim 10^{15}
\msun$ is $\sim\!\!10^{63}-10^{64}\ergs$.  As two clusters of galaxies merge,
the infall velocities can exceed the sound speed of the ICM.  As a result, a
shock front will form at the interaction boundary between the clusters.
First-order Fermi acceleration at the shock front produces a population of
nonthermal, relativistic particles.  Relativistic electrons are detected from
their synchrotron radio emission or from $\gamma$-rays due to
Compton-scattered cosmic microwave radiation. Nonthermal protons are detected
through $\gamma$ rays emitted from secondaries formed in nuclear production
processes, including the $\pi^0$ decay signature at 70\mev.

Rich clusters contain a hot and tenuous ionized intracluster medium (ICM),
with observed temperatures $T_X \approx 5-10\kev$, sound speeds
$\sim\!\!1000\kms$, and thermal bremsstrahlung luminosities $L_{X} \sim
10^{45}\ergss$ between 2 and 10\kev.  In addition to the luminous thermal
component present in these clusters, there is a growing body of evidence
supporting the presence of nonthermal distributions of particles in cluster
mergers \citep{eilek:99,feretti:00}.  Deep radio observations of clusters of
galaxies indicate the presence of extended diffuse emission not easily
associated with an optical counterpart.  These diffuse radio features are
commonly classified as either {\em radio halos}, or {\em radio relics} (which
are also called {\em periphery halos}).  Radio halos mimic the observed X-ray
profiles and are characterized by their central location in the cluster and by
a highly disorganized magnetic field, and are generally thought to be a
consequence of a merger event \citep{feretti:00}.  Radio relics, usually found
on the periphery of the cluster, are characterized by highly organized
magnetic fields and often display filamentary structures.  We focus here on
radio relics, which are thought to result from synchrotron emission emitted by
electrons directly accelerated at shock fronts. By contrast, radio halos might
be due to reacceleration of relic electrons by magnetic turbulence or enhanced
magnetic turbulence arising from the cluster merger or motions of the galaxies
\citep{brunetti:01,ohno:02}.

The first radio halo in a cluster of galaxies was detected from the Coma
Cluster \citep{large:59}.  Many halos and relics have since been found in a
number of other clusters, including A754 \citep{kassim:01}, A2256
\citep{berrington:02}, and others (see \citet{govoni:01,slee:01} for recent
observations). Typical radio powers from the radio halos and relics are at the
level of $10^{40}$-$10^{42}\ergss$ \citep{giovannini:00}.  These clusters show
a favorable correlation \citep{kassim:01,berrington:02} between the existence
of diffuse radio features and recent or on-going merger activity.  Detection
of these features is largely due to sensitivity limitations; the number of
detected diffuse radio features has increased with the improvement in
sensitivity of radio telescopes.

The synchrotron evidence for nonthermal electrons in clusters of galaxies
indicates that an acceleration mechanism must exist in the environment of the
host cluster with sufficient power to produce the observed emission.  Several
mechanisms have been proposed to explain the correlation of the radio halo and
relic features and the recent or on-going merger activity.  These mechanisms
often require the presence of shock waves or magnetic turbulence to accelerate
particles via the Fermi acceleration process
\citep{schlickeiser:87,tribble:93,kang:97,ensslin:98b,blasi:00,miniati:01}.
Other theories include adiabatic compression of fossil radio plasma by a
cluster-merger shock wave \citep{ensslin:00}.

Optical surveys show that approximately $30$-40\% of clusters of galaxies
display evidence for the presence of substructure \citep[and
others]{forman:81,gellers:82}.  This internal structure is often interpreted
as a subset of galaxies merging with a larger cluster of galaxies.  These
internal structures indicate velocity differences near or greater than the
expected sound speed of the intracluster medium (ICM).  Observed velocities
typically range from $\sim 1000$ to 3000\kms\, which are consistent with
values expected from parabolic orbits \citep{oegerle:94}.

Emission from nonthermal particles will also appear at EUV and hard X-ray
energies as a power-law excess.  While the EUV emission seen in the Coma
cluster \citep{lieu:99a}, and possibly also A2199 and A1795 \citep{lieu:99b}
may have a cool thermal origin, it is unlikely because of the extreme mass
requirements of cool gas. It is more likely that the EUV emission has a
nonthermal origin \citep{hwang:97,ensslin:98a,mittaz:98,bc99,av00}.  Excess
hard X-ray (HXR) emission has been reported from the clusters A1656
\citep{fusco-femiano:99,rephaeli:99}, A2256 \citep{fusco-femiano:00}, A3667
\citep{fusco-femiano:01} and possibly A2199 \citep{kaastra:99}.  Estimated
luminosity of the HXR emission are $\approx\!\!10^{43}\ergss$.  Thermal
origins of this HXR excess emission would require unrealistic temperatures
greater than $40\kev$, and so is thought to be caused by the presence of
relativistic nonthermal electrons\citep{fusco-femiano:99}.

Numerical models of merging clusters of galaxies \citep[and
others]{ricker:98,takizawa:00,ricker:01,miniati:01} have treated the
development of shocks as a result of the cluster merging process.  Given the
presence of thermal ionized particles in the vicinity of these shocks and a
cluster magnetic field, electrons and ions will be accelerated via the
first-order Fermi process \citep{bell:87,blandford:87}.  Numerous attempts
have been made to model the emissions from nonthermal particles
\citep{colafrancesco:98,fujita:01,petrosian:01,miniati:01} produced by these
cluster merger shocks, but this study differs from previous attempts in that
we accurately model the diffusion of particles in energy space due to Coulomb
interactions, and allow for a variable injection rate that depends on the
environment local to the shock front.  Recent studies by \citet{liang:02} have
addressed the balance of Coulomb losses and diffusion in energy space, but do
not address the injection and diffusion of particles in energy space along
with a variable source function of nonthermal particles.  In our treatment, we
follow particle energies up to $10^{19}\ev$, and calculate the bremsstrahlung,
Compton, synchrotron, and $\pi_{0}$ $\gamma$-rays from p-p collisions.
Because the shock front lifetime is a significant fraction of the age of the
universe, we also follow the changing environment due, for example, to the
changing CMB energy density.  In addition we accurately model nonthermal
electrons and protons up to $\sim\!\!10^{21}\ev$, though we find that
limitations on particle acceleration make it difficult to produce protons
above $\sim 10^{19}\ev$ in cluster merger shocks.  From the particle
distributions, we calculate nonthermal photon spectra for energies up to
$\sim\!\!10^{7}\gev$.  We also include the effects of secondary production on
the nonthermal photon spectra.

The physical processes and the temporal evolution of the particles are
described in section \ref{sec:models}.  The results of the simulations are
presented in section \ref{sec:results}.  A comparison of the photon spectra
with observed clusters, and the potential of detecting these shocks with
space-based satellite observatories is discussed in section
\ref{sec:discussion}.

\section{Models}
\label{sec:models}

We have developed a code to calculate the time-dependent particle distribution
functions evolving through radiative losses for electrons and protons
accelerated by the first-order Fermi process at the cluster merger shock.  The
code was originally adapted from a supernova remnant code \citep{sturner:97},
and applied to the specific case of the cluster merger scenario.  We will
break up the discussion of the problem in to four subsections: cluster merger
dynamics, nonthermal particle production, temporal evolution, and the
production of the photons.

\subsection{Cluster Merger Dynamics}
\label{sub:cluster_merger_dynamics}

\subsubsection{Evolution of the Merger Event}
\label{subsub:evolution_of_merger}

Two clusters are assumed to be the dominant gravitating masses in the local
region, and variations of the gravitational field due to surrounding masses
are assumed to be negligible. A cluster accretes another cluster which falls
from a distance $d$.  This distance can be approximated by the maximum
separation of two point masses which move apart beginning at the moment of the
big bang, follow an elliptical orbit, and merge at time $t_m$.  Given this
assumption, a cluster of mass $M_{1}=10^{15}M_{15}\msun$ accretes a subcluster
of mass $M_{2}=10^{15}m_{15}\msun$ at time {\em $t_{m}$} as approximated by
Kepler's third law.  Consequently, a cluster that merges at a time $t_{m}$
fell from a maximum separation approximated by
\begin{equation}
        \begin{array}{rcl}
        d & \approx & \left[ {\frac{2 G (M_1 + M_2) t_{m}^2}{\pi^2}}
        \right]^{\frac{1}{3}} \\ 
        d & \approx & 4.5 \left( \frac{M_1 + M_2}{10^{15} \msun} 
        \right)^{\frac{1}{3}} \left(\frac{t_{m}}{10^{10} yrs} \right)
        ^{\frac{2}{3}}\mpc
        \end{array}
\label{eqn:turn_around_distance}
\end{equation}
\citep{ricker:01}.  This equation allows us to calculate the magnitude of the
energy pool available for particle acceleration.  The total orbital
energy ${\cal E}$ of the merging system, assuming a zero
impact parameter, is given by
\begin{equation}
\begin{array}{rcl} 
        {\cal E} & \approx & -\frac{M_{1}M_{2}}{2^{1/3}
        (M_{1} + M_{2})^{1/3}} \left( \frac{\pi G}{t_{m}} \right)^{2/3}\\
        {\cal E} & \approx & -1.9\times 10^{64}\ \frac{M_{15} m_{15}}{(M_{15} +
        m_{15})^{1/3}} \left( \frac{t_{m}}{10^{10} yrs} \right)^{-2/3}
        \mbox{ergs}. 
\end{array}
\label{eqn:total_energy}
\end{equation}
We use the terminology that orbits are bound if the total energy ${\cal E} <
0$; otherwise, they are unbound.  It is customary to use the Keplerian analogy
of elliptical to represent orbits where ${\cal E} < 0$, and parabolic for
orbits where ${\cal E} = 0$.

To approximate the merger velocity of the system, we relax the assumption that
the accreting cluster is a point mass.  Instead, we assume the ICM number
density of the accreting cluster follows the spherically-symmetric, isothermal
beta model given by
\begin{equation} 
n_{\rm ICM}(r) = n_{0} \left[ 1 + \left(\frac{r}{r_{c}}\right)^{2}
\right]^{-\frac{3\beta}{2}}.
\label{eqn:number_density}
\end{equation}
The quantity $n_{0}$ is the central number density of the ICM, and $r_{c}$ is
the core radius, which characterizes the spatial scale over which the density
changes from a roughly constant value at $r \lesssim r_c$ to a power-law
density dependence at $r \gtrsim r_c$.  The gravitational dynamics are
dominated by the dark matter distribution of the more massive cluster.
Assuming that the dark matter distribution follows a similar profile as the
ICM, and that the merging cluster is a point mass, we calculate the cluster
merger velocity and radial separation by solving the differential equation
\begin{equation}
m_{r} \frac{d^{2} r_{12}}{dt^{2}} = -M_{2}
\frac{d\Phi[M_{1}(r_{12}),r_{12}]}{dr_{12}}.
\label{eqn:merger_velocity}
\end{equation}
The radius $r_{12}$ is the separation of the center of masses of the two
clusters, $M_{1}(r_{12})$ is the mass interior to the subcluster at radius
$r_{12}$, $M_{2}$ is the mass of the merging subcluster, and $m_{r} =
\frac{M_{1} M_{2}}{M_{1} + M_{2}}$ is the reduced mass.  The gravitational
potential $\Phi(M_{1}, r_{12})$ is defined by
\begin{equation}
\Phi(M, r) = -\frac{G M(r)}{r}\;.
\end{equation}
The mass interior is defined as
\begin{equation}
M(r) = 4\pi \xi_{m} \int_{0}^{r} dx ~ x^{2} n_{\rm ICM}(x)
\end{equation}
where $\xi_{m}$ is a normalization constant chosen so that $M_{1}(r_{12})$ is
equivalent to the total mass of the dominant cluster at its maximum radius
$R_{1}$.  Obviously, the potential satisfies the requirement
$\Phi(M,r\rightarrow\infty) = 0$. Initial conditions $(t=0)$ are found by a
conservation of energy argument where the initial radius and infall velocity
are given by
$$r_{12}(t=0) = R_{1} + R_{2}$$
\begin{equation}
        v_{0}(t=0) = \sqrt{\frac{2} {m_{r}} \left[{\cal E}-M_{2} \Phi(M_{1},
        r_{12})\right]}\;,
\label{eqn:initial_merger_velocity}
\end{equation}
where $R_2$ is the maximum radius of cluster $M_2$.

\begin{figure}
\begin{center}
\leavevmode
\hbox{%
\epsfxsize=6.5in
\plotone{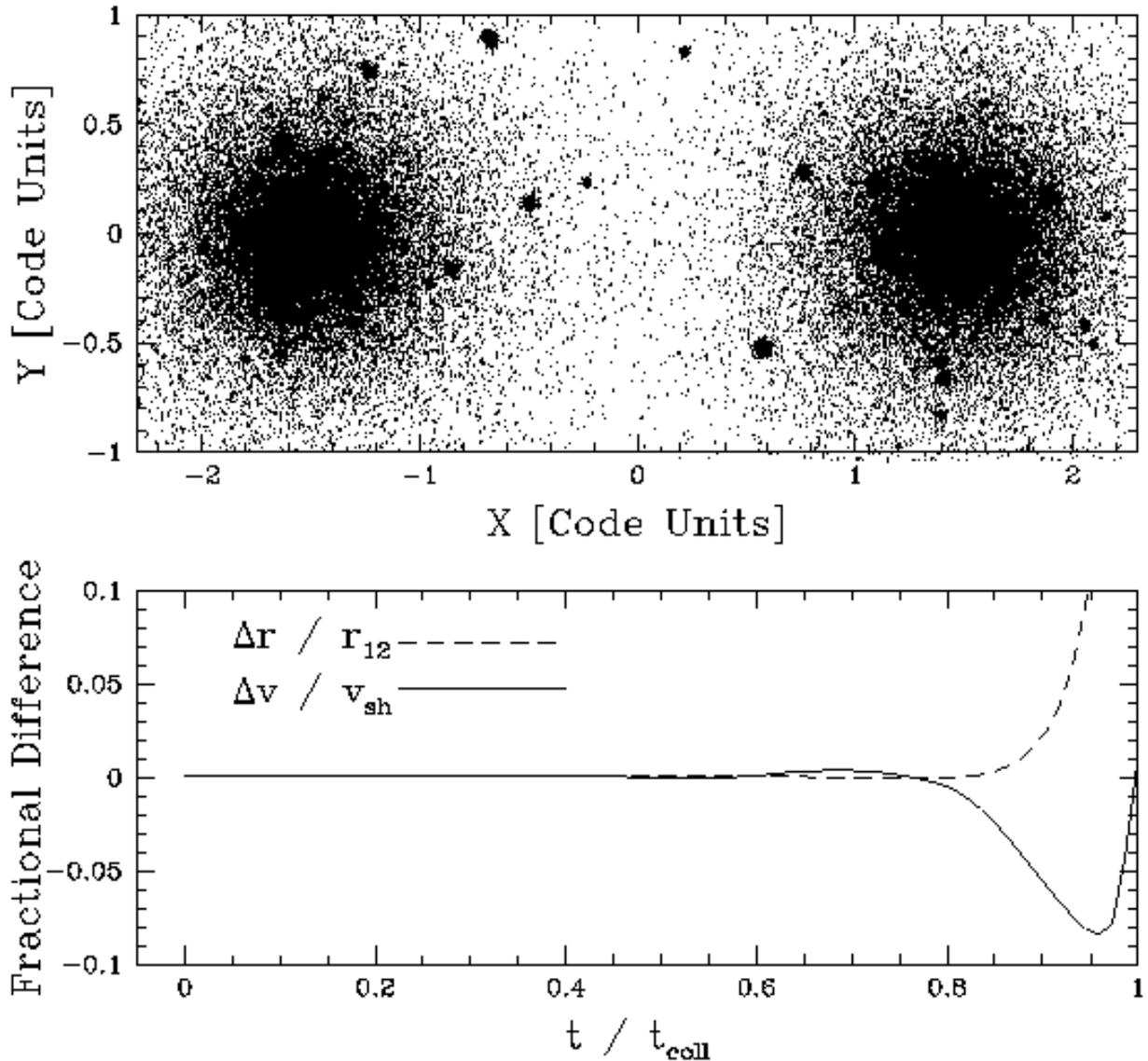}}
\caption{Comparison between the semi-analytic model used in subsequent
calculations and results of an N-body simulation. The initial configuration of
an N-body simulation of a cluster merger between two equal mass clusters is
shown in the top panel. The relative differences $\Delta r$ and $\Delta v$
between the semi-analytic and numerical calculations of the center-of-mass
distance and the infall velocity of the two clusters are shown in the lower
panel.  \label{fig:gravity_sim}}
\end{center}
\end{figure}

In order to test the accuracy of this approach to the dynamics of cluster
mergers, we compare the semi-analytic values of $r_{12}$ and $v_{0}$ with the
results of an N-body simulation of a two-cluster merger model
\citep{berrington:00}.  For this comparison, we assume that the two clusters
have equal masses. This is the worst case, because when one cluster is smaller
and less massive than the other, it is better described as a point source. In
the N-body simulation, each cluster is approximated by $N = 80,000$ equal-mass
particles, with 50\% of the particles assigned to galaxies in each cluster and
the remaining 50\% assigned to a non-luminous dark matter which follows a
King-model distribution with $W_{0}=6.25$ \citep{king:66}. The kinetic energy
of the particles in each cluster is assumed to be in virial equilibrium with
respect to the total gravitational potential energy prior to the interaction.
Each cluster consists of 50 galaxies with masses that are selected from a
Schechter luminosity function with a power-law index of $-1.25$. The top panel
of Fig.\ \ref{fig:gravity_sim} shows the initial configuration of the merger
event, and the lower panel gives the relative difference between the centers
of mass calculated in the N-body simulation and semi-analytically.

The calculation ends at the collision time $t_{\rm coll}$, which is defined as
the time when the centers of mass of the two clusters coincide. As can be
seen, the semi-analytic value of the $v_{0}$ is accurate to within 10\% at
times preceding $t_{\rm coll}$. After $t_{\rm coll}$, the forward shock
velocity, $v_{1}$, can no longer be approximated by equation
\ref{eqn:initial_merger_velocity} because the shock decouples from the
gravitational infall of the merging cluster, and is assumed to be constant as
the shock propagates outward from the cluster center.

In this paper, we model the cluster merger shocks rather than the accretion
shocks.  An infalling cluster is assumed to be a virialized, quasi-spherical
distribution of galaxies, ICM, and dark matter.  A cluster merger shock forms
when the cluster merger velocity exceeds the sound speed of the ICM.  In
contrast, an accretion shock is the result of a spherically symmetric infall
of matter from the surrounding volume onto a cluster.  In the context of the
previous work by \citet{fujita:01}, our definition of merger events includes
not only their definition of a cluster merger event($M_2 > 0.6 M_1$), but also
their definition of semi-merger events ($0.6 M_1 > M_2 > 0.1 M_1$).  In their
calculations, accretion events can account for only $\approx 10\%$ of the
nonthermal photon luminosity.  Although each cluster of galaxies is at the
center of an accretion flow, we do not consider nonthermal particle production
from accretion shocks because their nonthermal photon production is negligible
in comparison to cluster merger events.  Note that supersonic cluster merger
shocks are less likely to form at high redshift for two clusters of galaxies
with given masses. This is because the relative velocities of the merging
clusters will be associated with orbits with smaller separations at earlier
times, and will therefore have smaller relative velocities.  In reality, the
mean cluster masses are smaller at early times.  These lower mass clusters
have a lower virial temperature.  If this is considered in the calculation of
the strengths of the shocks seen in cluster merger events, it is possible that
stronger shocks may be seen at higher redshifts.

\subsubsection{The Cluster Environment}
\label{subsub:cluster_environment}

Shock formation and nonthermal particle production depends on the properties
of ICM, which are largely determined by the mass of the cluster and
cosmological epoch.  Below we describe the equations used to describe the
properties of the cluster environment.

X-ray observations of galaxy clusters show that these clusters are permeated
by a hot ICM of temperatures in the range of 5\kev-12\kev.  At these
temperatures the gas is well described by an ideal gas.  The sound speed of an
ideal gas in terms of the gas temperature $kT_{X}$ is simply
\begin{equation}
c_{s}(T_{X}) = \sqrt{\frac{\Gamma k T_{X}}{\mu m_{p}}} = 1265
\mu^{-\frac{1}{2}} \left( \frac{kT_{X}}{10 \kev} \right)^{\frac{1}{2}} \kmsec,
\label{eqn:sound_speed}
\end{equation}
where $\Gamma$ represents the ratio of specific heats which we take to be
equal to $5/3$, $\mu$ is the mean atomic mass, and $m_{p}$ is the proton mass.

The temperature of the ICM depends on both the mass of the cluster and
redshift $z$.  In order to calculate the cluster X-ray luminosity which gives
the energy of X rays that can be Compton scattered, we first calculate the
expected temperature of the ICM gas.  We use the cluster M-T relationship from
\citet{bryan:98}, given by
\begin{equation}
kT_{X}(z) = 1.39 f_{T_{X}} \left( \frac{M_{1}}{10^{15} \msun}
\right)^{\frac{2}{3}} (\Delta_{c} h^{2} E^{2}(z))^{\frac{1}{3}} \kev
\label{eqn:cluster_temperature}
\end{equation}
where $M$ is the mass of the cluster, $h$ is the parameter defined by $H =
100~ h~ \kmsec \mpc^{-1}$, the function $E^{2}(z) = \Omega_{0} (1 + z)^{3} +
\Omega_{R} (1 + z)^{2} + \Omega_{\Lambda}$, and $f_{T_{X}}$ is a normalization
constant taken to be $0.8$.  We have made use of the following standard
definitions:
\begin{equation}
\Omega_{0} = \frac{8 \pi G \rho_{c}}{3 H_{0}^{2}};\hspace{.3in} \Omega_{R} =
(H_{0} R)^{-2};\hspace{.3in} \Omega_{\Lambda} = \frac{\Lambda}{3 H_{0}^{2}}
\end{equation}
with \H0 as the current measured Hubble constant, $\rho_{c}$ is the critical
density required to close the universe, $\Lambda$ is the cosmological
constant, and $R$ is the current curvature radius of the universe.  We force
the constraint $\Omega_{0} + \Omega_{R} + \Omega_{\Lambda} = 1$ as mandated by
an inflationary cosmology.  For all models in this paper, we set $\Omega_{R} =
0$ for a flat universe.  The function $\Delta_{c}$ is the critical density
factor and marks the point at which a region of overdensity makes the
transition from linear density growth behavior to a nonlinear density growth
regime.  This marks the birth of a cluster or the beginning of the initial
collapse of a density perturbation out of the Hubble flow.  \citet{bryan:98}
calculate it to be
\begin{equation}
\Delta_{c} = \left\{ 
        \begin{array}{ll}
        18 \pi^{2} & \mbox{for}~ \Omega_{0} = 1 \\
        18 \pi^{2} + 82 x - 39 x^{2} & \mbox{for}~ \Omega_{R} = 0\\
        18 \pi^{2} + 60 x - 32 x^{2} & \mbox{for}~ \Omega_{\Lambda} = 0
        \end{array},
\right.
\label{eqn:critical_density}
\end{equation}
where $x = \Omega(z) - 1$, and $\Omega(z) = \Omega_{0} (1 + z)^{3} /
E^{2}(z)$.  These relations are accurate to 1\% in the range $\Omega(z) =
0.1$-$1.0$.

A number of radiation fields provide seed photons for the Compton scattering
by the nonthermal electrons present in the cluster.  For the X-ray radiation
field, we calculate the X-ray luminosity of a cluster of mass $M$ by the
observed relation \citep{arnaud:99}
\begin{equation}
L_{X}(T_{X}) = 2.88 h^{-2} \times 10^{44} \left( \frac{kT_{X}(z)}{6 \kev}
\right)^{2.88} \ergss\;,
\label{eqn:cluster_luminosity}
\end{equation}
where $kT_{X}$ is defined by equation (\ref{eqn:cluster_temperature}).
Assuming the X-ray luminosity is uniformly distributed within a spherical
cluster of radius $R$ and an escape time of $\sim\!\!R/c$, the mean energy
density of the X-ray photons is
\begin{equation}
U_{X}(T_{X}) \approx 1.5 h^{2} \times 10^{-10} \left( \frac{R_{1}}{1
\mpc} \right)^{-2} \left( \frac{kT_{X}(z)}{6 \kev} \right)^{2.88} \mevcc.
\label{eqn:xray_energy_density}
\end{equation}
The thermal X-ray bremsstrahlung is approximated by a blackbody of temperature
$T_{X}$ in our calculations of Compton scattering. This is not a crucial
approximation, because the X-ray energy density is $\sim 2$-3 orders of
magnitude less than the CMB energy density, and the importance of
Compton-scattered X-ray photons is furthermore reduced due to the
Klein-Nishina effects.

A second component that contributes to the mean photon energy density of the
ICM is the stellar photon field.  We assume that the galaxy luminosity
function is well approximated by the Schechter luminosity distribution:
\begin{equation}
n_{\rm gal}(L)\ = \frac{\xi_{\star}}{L^{\star}} \left(
\frac{L}{L^{\star}} \right)^{-\alpha} \exp \left[-\frac{L}{L^{\star}} \right]
\label{eqn:schechter_luminosity_function}
\end{equation}
where $n_{\rm gal}(L)\ dL$ is the differential luminosity function and is
defined as the number of galaxies within the luminosity range $L$ and $ L +
dL$.  The parameter $L^{\star}$ is the ``characteristic luminosity'' of a
galaxy which marks the transition between the faint-end power law of slope
$-\alpha$ and the high-end exponential cutoff.  The constant $\xi_{\star}$
normalizes the Schechter function to $N_{gal}$ when integrated over all
luminosities.  We can use equation \ref{eqn:schechter_luminosity_function} to
calculate the total stellar luminosity of
\begin{equation}
L_{\rm star} = \int_0^{\infty} dL~L~n_{\rm gal}(L) = \xi_{\star}
\Gamma[2 - \alpha] L^{\star}
\label{eqn:total_stellar_luminosity}
\end{equation}
where $\Gamma[2 - \alpha]$ is the gamma function.  Typical values for
$\xi_{\star}$, $L^{\star}$, and $\alpha$ are given, e.g., in
\citet{paolillo:01}, and our adopted values are $\alpha = 1.1$, a value of
$L^\star\cong 7\times 10^{43}$ ergs s$^{-1}$ corresponding to an absolute
magnitude $M^{\star} = -20.5$ in $B_J$ \citet{colless:89}, and $\xi_{\star} =
100$.

Using the same approach as for the X-ray photon field, the estimated mean
stellar radiation field becomes
\begin{equation}
U_{\rm star} \approx 1.4 h^{2} \times 10^{-9} \left(
\frac{R_{1}}{1 \mpc} \right)^{-2}\mevcc .
\label{eqn:stellar_energy_density}
\end{equation}
The spectral distribution of elliptical galaxies is well approximated by a
K3III star \citep{pierce:02}.  Since higher density regions in rich galaxy
clusters are dominated by elliptical galaxies, we will assume that the stellar
radiation energy distribution can be approximated by a blackbody spectrum
resembling that of a K3III star at temperature $T_{\rm star} \approx 4000$ K.

The energy density of the CMB is 
\begin{equation}
U_{\rm CMB} = 2.5\times 10^{-7} (1 + z)^{4} \mevcc.
\label{eqn:CMB_energy_density}
\end{equation}
The CMB is the dominant contributor to total cluster photon energy density by
approximately 3 orders of magnitude.  The total cluster radiation field is
simply the sum of the equations (\ref{eqn:xray_energy_density}),
(\ref{eqn:stellar_energy_density}), and (\ref{eqn:CMB_energy_density}).

\subsubsection{Cluster Redshift}

The cluster environment described in section \ref{subsub:cluster_environment}
depends on the redshift of the cluster.  The redshift evolution of the cluster
depends on the chosen cosmological model.  We have chosen to uniquely specify
the cosmological model by the following parameters: critical mass fraction
$\Omega_{0}$, the curvature $\Omega_{R}$, the dark energy $\Omega_{\Lambda}$,
and the Hubble constant \H0.  For this paper we will consider the model
$(\Omega_{0}, \Omega_{R}, \Omega_{\Lambda}) = (0.3, 0.0, 0.7)$.

The observed redshift of a cluster is determined from the time since the Big
Bang by the cosmological equation \citep{peebles:93}
\begin{equation} 
\frac{dz}{dt} = -H_{0} (1+z) E(z) 
\label{equ:redshift_relation}
\end{equation}
where $E^{2}(z)$ is the dimensionless parameter defined following equation
(\ref{eqn:cluster_temperature}).  Typical propagation times for a shock front
to traverse a cluster are $\sim\!\!10^{9}\yrs$.  This is a significant time
span in that the evolution of the cluster environment due to cosmic expansion
must be considered.  Cluster merger events can be initiated at any time
throughout age of the universe, and evolving the cluster environment is
necessary to accurately calculate the nonthermal photon and particle spectra.

\subsection{Nonthermal Particle Production}
\label{sub:nonthermal_particle_production}

Henceforth we denote particle kinetic energies by $K_{e,p} =
m_{e,p}(\gamma_{e,p} - 1) c^{2}$, total energies by $E_{e,p}$, and
dimensionless total energies by $\epsilon_{i}$, with the subscripts $e$ and
$p$ referring to electrons and protons, respectively.  First-order Fermi
acceleration at a cluster merger shock is approximated by injecting power-law
momentum spectra for the electron and protons.  The total particle injection
function in terms of kinetic energy is given by
\begin{equation}
Q_{e,p}(K_{e,p},t) = Q_{e,p}^{0} [K_{e,p} (K_{e,p} + 2 m_{e,p}
c^{2})]^{-\frac{s(t) + 1}{2}} (K_{e,p} + m_{e,p} c^2) \exp\left[ -
\frac{K_{e,p}}{K_{\rm max}(t)} \right],
\label{eqn:power-law}
\end{equation}
where $s(t)$ is the injection index, and $K_{\rm max}$ is the maximum particle
energy determined by three conditions: the available time to accelerate to a
given energy since the beginning of the merger event; the requirement that the
particle Larmor radius is smaller than the size scale of the system; and the
condition that the energy-gain rate through first-order Fermi acceleration is
larger than the energy-loss rate due to synchrotron and Compton processes.
The duration of particle injection is determined by the dynamics of the
system, and the shock speeds are determined by equation
(\ref{eqn:merger_velocity}).  The shocks terminate at the time $t_{\rm acc,1}$
for the forward shock and $t_{\rm acc,2}$ for the reverse shock when the shock
propagates to the outer part of the cluster.  The time $t_{\rm acc}$
represents the greater value of two acceleration times $t_{\rm acc,1}$ and
$t_{\rm acc,2}$.  The constant $Q_{e,p}^{0}$ normalizes the injected particle
spectrum over the volume swept out by the shock front, and is determined by
\begin{equation}
\int_{K_{\rm min}}^{K_{\rm max}} dK_{e,p} ~ K_{e,p} Q_{e,p}(K_{e,p},t) =
\frac{\eta_{e,p}}{2} A \; \eta^{e}_{\rm He} m_{p} v_{1}(t)^{3} \langle n_{\rm
ICM}(t) \rangle\;,
\label{eqn:normalization}
\end{equation}
where $A$ is the area of the shock front, $\eta^{e}_{\rm He} \cong 1.2$ is an
enhancement factor due to the presence of ions heavier than Hydrogen, and
$\langle n_{\rm ICM}(t) \rangle$ is the number density of the gas averaged
over the area of the shock front at time $t$.  We assume a constant efficiency
factor $\eta_{e,p} = 5\%$ for both electrons and protons, though $\eta_{e,p}$
would depend on Mach number through the uncertain wave generation processes at
the shock front.\footnote{A quantitative prescription for the shock efficiency
  as a function of the injection index is given by \citet{miniati:01}; see
  also \citet{kes03} on this issue.} The particle injection index, $s(t)$, is
determined by the shock properties described in \S
\ref{subsub:particle_source_function}.  Typical values for the particle
injection index are $s \approx 2.5$.

The temporally evolving particle spectrum is determined by solving the
Fokker-Planck equation in energy space for a spatially homogeneous ICM, given
by
$$\frac{\partial N(K,t)}{\partial t} = \frac{1}{2} \frac{\partial^{2}}{\partial
K^{2}} \left[D(K,t)~N(K,t)\right] - \frac{\partial}{\partial K}
\left[\left(\dot{K}_{\rm tot}(K,t) - A_{\rm tot}(K,t)\right) N(K,t)\right] $$
\begin{equation} -
\sum_{i=pp,p\gamma,d}\frac{N(K,t)}{\tau_{i}(K,t)} + Q(K,t).
\label{eqn:fokker_planck}
\end{equation}
The quantity $\dot{K}_{\rm tot}$ represents the total kinetic-energy loss
rate, and $A_{\rm tot}$ represents the total energy gain rate.  Electrons
experience energy losses from synchrotron, bremsstrahlung, Compton, and
Coulomb processes.  The total energy loss rate for protons is due to the
effects of Coulomb processes.  In addition, protons experience catastrophic
energy losses due to proton-proton collisions ($i=pp$), proton-$\gamma$
collisions ($i=p\gamma$), and diffusive escape ($i=d$) on the time scale
$\tau_{i}(K_{p},t)$.  The spectra of the secondary electrons are calculated
from the pion-decay products (see Appendix
\ref{appsub:secondary_electron_production}), and are subject to the same
energy losses as the primary electrons.  The calculation of the energy loss
rates accounted for by the quantity $\dot{K}_{\rm tot}$ for both electrons and
protons is presented in detail in Appendix
\ref{app:particle_energy_loss_rates}.  The nonthermal energy spectra resulting
from the temporal evolution of equation (\ref{eqn:fokker_planck}) will produce
a nonthermal photon spectra.  The method used to calculate the synchrotron,
bremsstrahlung, Compton, and $\pi_{0}$ $\gamma$-ray radiation components to
the photon energy spectra is presented in detail in Appendix
\ref{app:photon_production}.

The integration of the partial differential equation (\ref{eqn:fokker_planck})
poses a significant computational problem when considering the ranges of
energies modeled in this paper.  The problem stems from the diffusion that
occurs at low energies, and the large energy losses that occur at high
energies.  We found that equation (\ref{eqn:fokker_planck}), when solved with
a Crank-Nicholson implicit finite differencing method, is unstable to an
oscillation instability at the high and low energies due to the large values
that the convection derivative takes at both energy extremes.  To solve
equation (\ref{eqn:fokker_planck}), we settled on an implicit, upwind finite
differencing method.  This has the advantage of solving the oscillation
instability problem, though with the disadvantage of a decrease in numerical
accuracy for a given stepsize.

%
%

\subsubsection{Particle Source Function}
\label{subsub:particle_source_function}

Nonthermal particle distributions are injected into the ICM according to the
function given by equation (\ref{eqn:power-law}).  The normalization constant
$Q_{e,p}^{0}$ is calculated as described in \S
\ref{sub:nonthermal_particle_production}.  The particle distribution functions
have spectral index $s(t)$ and exponential cutoffs at kinetic energy $K_{\rm
max}$, which are determined by shock acceleration theory as described below.

Nonthermal particles gain energy through first-order Fermi shock acceleration
at the rate
\begin{equation}
\dot{K}_{sh}(t) = 100 \frac{B v^{2}_{8}(t)}{f r_{J}} \mevs
\label{eqn:particle_acceleration}
\end{equation}
\citep{sturner:97,reynolds:96}, where $v_8(t) = v_{1}(t)/(10^8$ cm s$^{-1}$)
is the shock velocity in units of $10^{3}$ \kmsec, $f$ is a factor giving the
mean free path of a particle in units of its gyroradius, and the quantity
$r_{J}$ accounts for the orientation of the shock front relative to the
magnetic field.  In our calculations, we set $f = 10$ and $r_J = 1$.

The first constraint that we impose on maximum particle energy is the
available time to accelerate particles to a given energy.  This constraint is
determined through the expression
\begin{equation}
K_{{\rm max},1} = \int^{t}_{0} dt~\dot{K}_{sh}(t)\;.
\label{eqn:maximum1_particle_energy}
\end{equation}
where $v_{1}(t)$ is found from equation (\ref{eqn:merger_velocity}), and $B $
is assumed to be constant throughout the cluster volume.  This constraint is
the most important limitation on proton energy for most of the lifetime of the
shock front.

Competition with energy losses sets a second constraint on maximum particle
energy.  By equating the electron energy gain rate, equation
(\ref{eqn:particle_acceleration}), with the sum of the synchrotron and Compton
energy losses that are the dominant electron energy-loss processes at high
energies, we obtain a second energy constraint for electrons, given by
\begin{equation}
K_{{\rm max},2} = {2.8 \times 10^{7} \;\frac{v_{8}(t)}{\sqrt{f r_J}}}\; \left[
\frac{B_{-7}} {(1+z)^{4} + 10^{-3} B^{2}_{-7}} \right]^{\frac{1}{2}} \mev.
\label{eqn:maximum2_electron_particle_energy}
\end{equation}
where $B_{-7}$ is the magnetic field in units of $10^{-7}$ G.  Considering
synchrotron losses only, the corresponding maximum energy for protons is is
\begin{equation}
K_{{\rm max},2(syn)} \cong 2 \times 10^{15} \frac{v_{8}(t)}{\sqrt{B_{-7}f
r_J}} \mev.
\label{eqn:maximum2_proton_particle_energy}
\end{equation}
For the parameters used in our calculations, proton energy losses due to
Compton and synchrotron processes are smaller than photomeson and photopair
losses with the CMB.  These latter processes could in principle constrain the
maximum proton energy \citep{berezinsky:88}, but are found to be negligible
for the parameters used in the calculations. This is because the time scale to
accelerate protons to ultra-high energies ($\gtrsim 10^{19}$ eV) requires
$\approx 3\times 10^{11} [fr_J/B_{-7} v_8^2(t)](K_p/10^{20}{\rm eV})$ yr,
which is longer than the Hubble time $t_{\rm H}$ and much longer than the time
scale over which the merger shocks persist in the merging clusters ($\approx
0.1 t_{\rm H}$). Although photomeson production does not pose a significant
limitation to particle acceleration, it is still treated as a catastrophic
loss in the calculation.

In order for particle acceleration to occur, the Larmor radius of a particle
must be smaller than the size scale of the system; otherwise, the particle
escapes the system before any significant energy gain can occur.  The third
constraint imposed on maximum particle energy is therefore obtained by
requiring that the particle Larmor radius be less than the size scale of the
shock front.  This constraint becomes, for both protons and electrons,
\begin{equation}
K_{{\rm max},3} = \frac{e B \lambda }{f} \cong  \frac{10^{13}}{f}\;
B_{-7} \;\left({\frac{\lambda}{\rm Mpc}}\right)\; \mev
\label{eqn:maximum3__particle_energy}
\end{equation}
where $\lambda = 2R_2$ and $e$ is the electron charge.  In all of our
calculations, the limitations imposed by $K_{{\rm max},1}$ and $K_{{\rm
max},2}$ are more restrictive for the electrons.  However, for protons
$K_{{\rm max},3}$ is the strongest limitation for the highest energy protons
at the later stages of the shock evolution.  The maximum energy $K_{max}(t)$
in equation (\ref{eqn:power-law}) is the minimum of the maximum energies
calculated in equations
(\ref{eqn:maximum1_particle_energy})--(\ref{eqn:maximum3__particle_energy}).
Note that the size scale limitation on maximum particle energy assumes that
the coherence length of the magnetic field is larger than the size scale of
the shock.

The particle index $s(t)$ is determined from the collision and sound speeds of
the system.  The collision speed $v_{1}(t)$ is obtained by solving equation
(\ref{eqn:merger_velocity}) with the initial radius of the interaction given
by the sum of the assumed radii of the clusters and the initial collision
velocity given by equation (\ref{eqn:initial_merger_velocity}).  We assume
that the subcluster is a point mass and the dominant cluster is described by
the isothermal beta model density profile given in equation
(\ref{eqn:number_density}).

\begin{figure}
\begin{center}
\leavevmode
\hbox{%
\epsfxsize=6.5in
\plotone{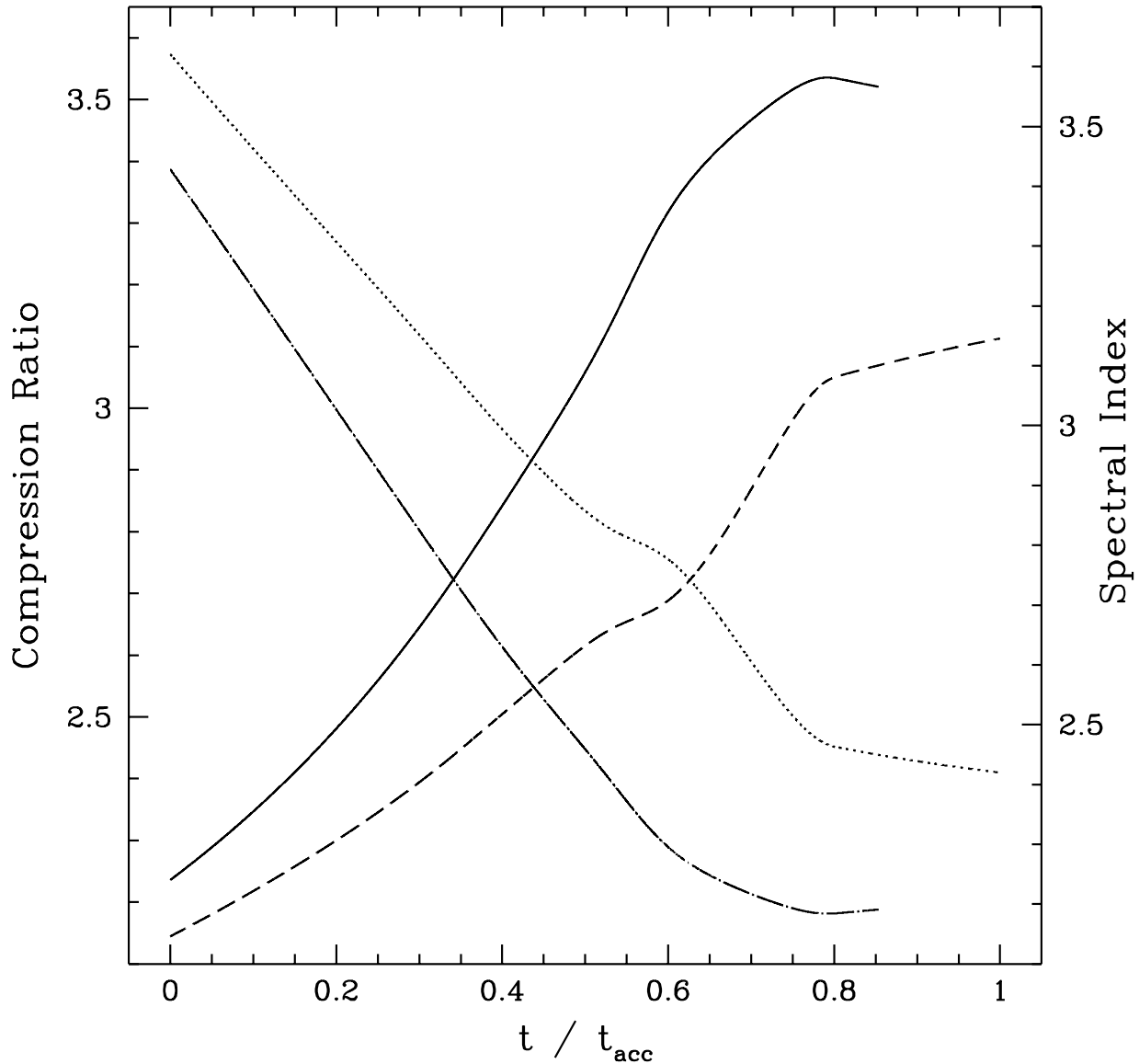}}
\caption{The compression ratio of the forward (dashed curve) and reverse
  (solid curve) shocks as calculated by the method described in \S
  \ref{subsub:particle_source_function}. The particle injection spectral index
  for the forward (dotted curve) and reverse (dot-dashed curve) shocks is
  associated with the compression ratio, and is calculated for a merger system
  consisting of a dominant cluster of mass $10^{15}\msun$ and a poor
  subcluster of mass $10^{14}\msun$.  Initial onset of the shock begins at a
  radius of $1.5\mpc$ from the cluster center with an initial redshift of $z_i
  = 0.1$. Particle acceleration occurs over a period $t_{\rm acc} = 1.1$ Gyr
  for these parameters, after which injection stops. Acceleration at the
  reverse shock ends after the reverse shock passes through the merging
  cluster.
\label{fig:compression_ratio}}
\end{center}
\end{figure}

The shock compression ratio is calculated from the equation
\begin{equation} 
{\cal C}(t) = \frac{\Gamma + 1}{\Gamma - 1 + 2 {\cal M}^{-2}(t)}
\label{eqn:compression_ratio}
\end{equation}
where ${\cal M}_{1}(t)=v_{1}(t)/c_s$ is the Mach number of the cluster merger
shock speed, as computed from equations (\ref{eqn:merger_velocity}),
(\ref{eqn:sound_speed}), and (\ref{eqn:cluster_temperature}). The ratio of
specific heats is represented by $\Gamma=5/3$ appropriate to a nonrelativistic
mono-atomic ideal gas.  In Figure \ref{fig:compression_ratio}, the time
evolution of the compression ratio of a shock front in a merging cluster with
masses $M = 10^{15}\msun$ and $m = 10^{14}\msun$ is graphed.  The radii of the
two systems are assumed to be $1.5\mpc$ and $0.75\mpc$, and the onset of the
merger is at a redshift of $z_i=0.1$.  From the shock compression ratio, we
calculate the injected particle index $s$ from
\begin{equation}
s(t) = \frac{{\cal C}(t) + 2}{{\cal C}(t) - 1}.
\label{eqn:injected_power_law_index}
\end{equation}
The calculated shock speeds of cluster merger shocks range from $\sim\!\!1.5$
to $3$ times the ICM sound speed, with power-law indices ranging from
$\sim\!5$ to $\sim\!2.3$. Figure \ref{fig:compression_ratio} also shows the
evolution of the typical power-law index as a function of time.

In the merger event between two clusters of galaxies, as described in \S
\ref{sub:cluster_merger_dynamics}, two shock fronts develop. The shock front
propagating into the dominant cluster, which we call the forward shock, has
Mach number ${\cal M}_1(t)$. Similarly, the shock front propagating into the
merging cluster has Mach number ${\cal M}_2(t)$.  Writing down the compression
ratio for the forward (${\cal C}_{1}(t)$) and reverse (${\cal C}_{2}(t)$)
shocks from equation (\ref{eqn:compression_ratio}), we get
\begin{equation} 
{\cal C}_{1}(t) = \frac{u_{1}(t)}{v} = \frac{4}{ 3 (1-{\cal M}_{1}^{-2}(t))}
\mbox{, and } {\cal C}_{2}(t) = \frac{u_{2}(t)}{\bar{v}} = \frac{4}{ 3
(1-{\cal M}_{2}^{-2}(t))},
\label{eqn:shock_velocity}
\end{equation}
where $u_{1}(t)$ and $u_{2}(t)$ are the velocities of the forward and reverse
shocks, respectively.  In addition, $v$ and $\bar{v}$ are the velocities of
the shocked fluid for the forward and reverse shocks, respectively.  Noting
that $\bar{v} = v_{0}(t) - v$ where $v_{0}$ is the infall velocity of the
merging cluster, and solving for the Mach number, we get
$${\cal M}_1(t) = \frac{2} {3} \frac{v}
{c_{1}} \left( 1 +
\sqrt{1 + \frac{9} {4}
\frac{c_{1}^{2}} {v^{2}} } \right) \;, $$
\noindent and
\begin{equation}
{\cal M}_2(t) = \frac{2} {3} \frac{v_{0}(t) - v} {c_{2}} \left( 1 +
    \sqrt{1 + \frac{9} {4} \frac{c_{2}^{2}} {(v_{0}(t) - v)^{2}} } \right)\;,
\label{eqn:forward_reverse_mach_number}
\end{equation}
where $c_{1}$ is the sound speed in the dominant cluster, $c_{2}$ is the sound
speed in the merging cluster, and the Mach number is defined to be ${\cal
M}_{1,2} = u_{1,2}/c_{1,2}$.  The values of $c_{1}$ and $c_{2}$ are calculated
from equation (\ref{eqn:sound_speed}) with temperatures given by equation
(\ref{eqn:cluster_temperature}) for clusters of mass $M_{1}$ and $M_{2}$.  The
value of $v$ is calculated by solving
\begin{equation}
\frac{\mu_{1}}{\mu_{2}} \frac{n_{1}(t)}{n_{2}(t)} = \frac{1 + 3{\cal
    M}_{1}^{-2}(t)}{1 + 3 {\cal M}_{2}^{-2}(t)} \left( \frac{v_{0}(t) - v}{v}
    \right)^{2}\;,
\label{eqn:relative_shock_velocity}
\end{equation}
where $n_{1}(t)$ and $n_{2}(t)$ are the ICM number densities in the dominant
cluster and merging cluster at the positions of the forward and reverse
shocks, respectively, and are calculated according to equation
(\ref{eqn:number_density}).  We take the mean mass per particle $\mu_1 = \mu_2
= 0.6 m_p$ in the two clusters.  Equation (\ref{eqn:relative_shock_velocity})
was derived from the shock jump conditions by equating the energy densities of
the shocked fluids at the contact discontinuity.  The compression ratios are
then calculated from equation (\ref{eqn:compression_ratio}).  These values are
plotted in Figure \ref{fig:compression_ratio}.

%
%

\section{Results}
\label{sec:results}

We present results of the numerical model described in \S \ref{sec:models} to
simulate the temporal evolution of nonthermal particles accelerated by shocks
formed in merging clusters of galaxies.  From these particle distributions, we
are able to investigate the expected nonthermal photon spectra for energies
ranging from $\sim\!\!10^{-9}$\ev\ to $\sim\!\!10^{19}$\ev\ that cover
astrophysically interesting frequencies.  At photon energies $\lesssim 4\times
10^{-8}$\ev\ ($\approx$ 10 MHz), ground-based radio observations are subject
to the ionospheric absorption cutoff.  At photon energies $\gtrsim 10^{12}$ eV
(depending precisely on source redshift) the universe becomes opaque to
$\gamma$ rays due to pair production attenuation on the diffuse infrared
radiation fields. The electron-positron pairs produced by the $\sim
10^{14}$-$\sim 10^{19}$ eV provide a secondary source of synchrotron and
Compton-scattered radiation on the Mpc galaxy-cluster scale, but this cascade
radiation is found to provide only a minor nonthermal synchrotron component to
the nonthermal UV/soft X-ray emission, and is not included.  To calculate
detectability of clusters with ground-based Cherenkov telescopes,
$\gamma\gamma$ absorption of the high-energy $\gamma$ rays must, however, be
considered.

\begin{figure}
\begin{center}
\leavevmode
\hbox{%
\epsfxsize=6.5in
\plotone{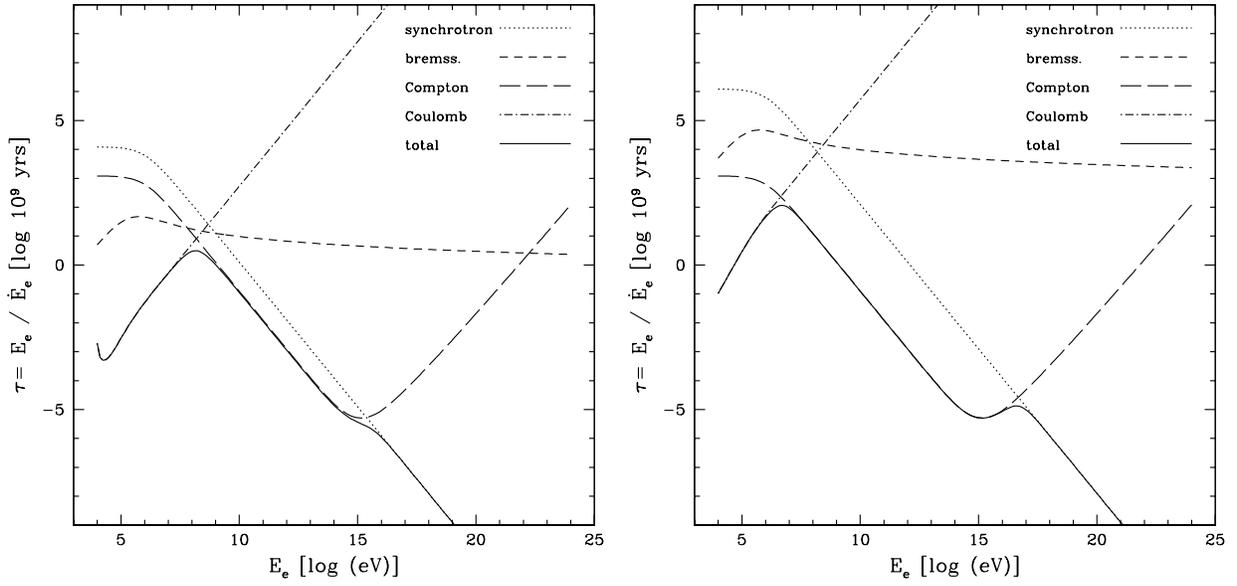}}
\caption{The instantaneous energy-loss timescale for electrons in a merging
  cluster scenario at redshift $z=0.0$.  In the left panel, the proton density
  $n = 10^{-3}\percc$ and $B = 1\mug$. Only the dominant CMB radiation field
  is considered.  In the right panel, $n=10^{-6}\percc$ and $B=0.1\mug$.
  \label{fig:electron_lifetime}}
\end{center}
\end{figure}

The energy-loss time scales of nonthermal electrons for various parameters
assumed in cluster environments are shown in Fig.\
\ref{fig:electron_lifetime}.  At low energies, Coulomb losses dominate.  A
maximum in the energy-loss time scale is found in the energy range from
$\approx 1\mev$ to several hundred $\mev$.  At higher energies, Compton and
synchrotron losses dominate. At the highest energies, the Compton loss rate
declines due to the onset of the Klein-Nishina reduction in the Compton cross
section, and synchrotron losses become the dominant loss mechanism (see also
\citet{sar99}). The electron loss-rate from double Compton scattering is
always much smaller than synchroton losses in our study, and can be safely
neglected (see \ref{appsub:electron_loss_rates}).

\begin{figure}
\begin{center}
\leavevmode
\hbox{%
\epsfxsize=6.5in
\plotone{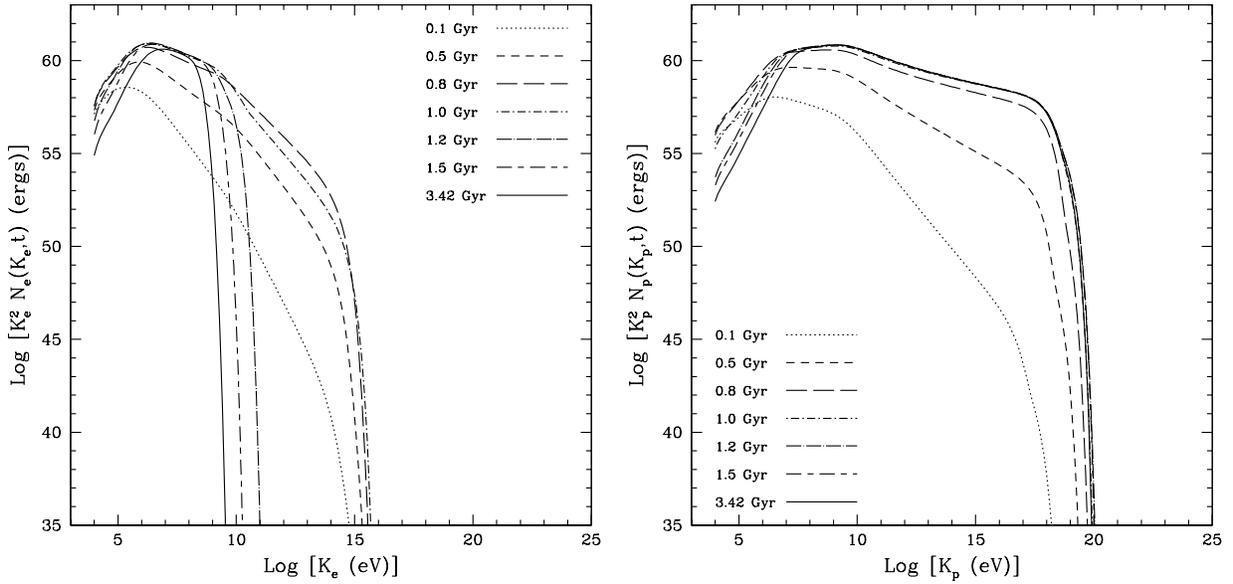}}
\caption{Primary electron ({\em left panel}) and proton ({\em right panel})
  energy spectra for a shock formed by a merging cluster using a standard
  parameter set with $M_1 = 10^{15}\msun$ and $M_2 = 10^{14}\msun$ and $B =
  1.0\;\mu$G.  The interaction begins at redshift $z_i = 0.3$ and is evolved
  to the present time ($z=0.0$; solid curve) with spectra shown at the various
  times given in the legend. The redshifts corresponding to the time delays of
  $0.1, 0.5, 0.8, 1.0, 1.2, 1.5$, and 3.42 Gyr are $z = 0.289, 0.248, 0.218,
  0.199, 0.180,0.153$ and 0.0, respectively.
 \label{fig:time_energy_spectra}}
\end{center}
\end{figure}

Figure \ref{fig:time_energy_spectra} shows the nonthermal particle energy
spectra at different times during the merger of a dominant cluster of mass
$M_1 = 10^{15} \msun$ and radius $R_1 = 1.5$\mpc\ with a subcluster of mass
$M_2 = 10^{14} \msun$ and a radius $R_2 = 0.75$\mpc, respectively.  The
quantities $M_{1}$, $M_{2}$, $R_{1}$, and $R_{2}$ are defined in \S
\ref{subsub:evolution_of_merger}.  The assumed core radius $r_c$ for the
dominant cluster is $250$\kpc.  The ICM density of the dominant cluster is
assumed to follow equation (\ref{eqn:number_density}) with a central number
density $n_{0} = 10^{-3}$\percc and $\beta = 0.75$. These parameters are
consistent with the $L_X$-$ T_X$ relation (\ref{eqn:cluster_luminosity}).  A
uniform magnetic field of strength $1$\mug\ was assumed throughout the
cluster.  The galaxy cluster environment and its evolution is calculated using
the cosmological parameters $(\Omega_{0}, \Omega_{R}, \Omega_{\Lambda}) =
(0.3, 0.0, 0.7)$.  All models assume a Hubble constant $\H0 = 70\kms\permpc$.
We also assume a Hydrogen to Helium number ratio of 10.

Irrespective of the specific parameters of the simulation, we see that at
energies $\lesssim 1\mev$, the electron and proton spectra display low-energy
cutoffs due to Coulomb diffusion and energy losses (the thermal particle
distributions are not shown). Electrons with energies greater than $\sim
1\gev$ lose a significant fraction of their energy on Gyr time scales. The
dominant energy-loss mechanism for these electrons is Compton scattering, as
Thomson losses on the CMB dominate synchrotron losses when $B \ll 3(1+z)^2$
\mug\ (see Fig.\ \ref{fig:electron_lifetime}).  The dominant energy-loss
mechanism for protons is secondary nuclear production, which operates on a
time scale of about $2 t_{\rm H}$ for an ambient density of $10^{-3}$
cm$^{-3}$. Consequently, the high-energy cutoff in the proton distribution
function results from the available time to accelerate particles and, at the
longest times, competition of particle acceleration with catastrophic
$p\gamma$ losses. Even though particles are injected with a range of power-law
indices, most of the injection power occurs when the spectral index is
hardest, yielding a proton spectrum that can be accurately described by a
single power law (compare Fig.\ \ref{fig:compression_ratio}) with a number
index of $\approx -2.5$ for the forward shock, and $\approx -2.2$ for the
reverse shock.

\begin{figure}
\begin{center}
\leavevmode
\hbox{%
\epsfxsize=6.5in
\plotone{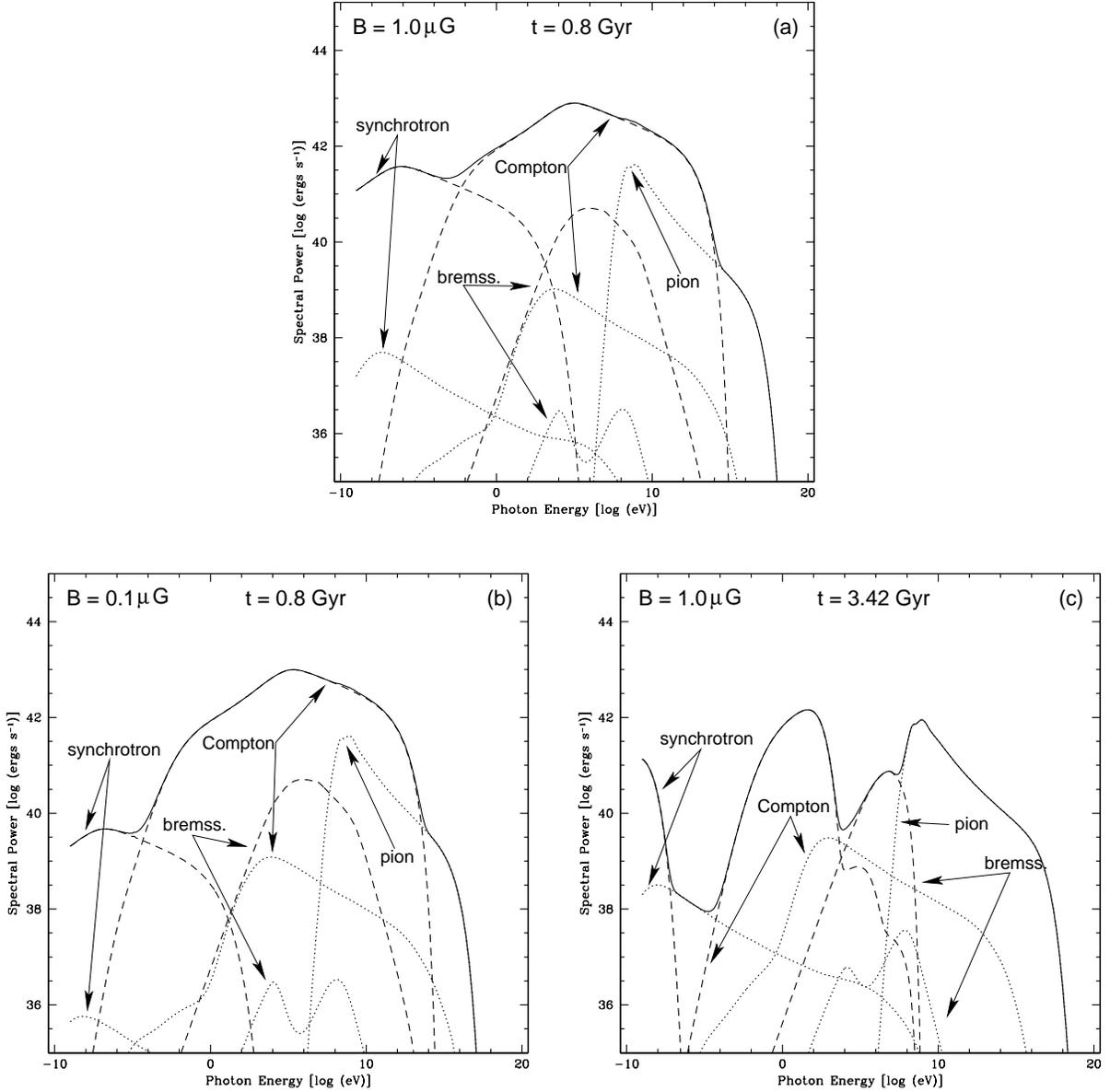}}
\caption{Separate radiation components produced by nonthermal particles
  accelerated by a shock initiated at $z_{i} = 0.3$ resulting from the merger
  between $10^{14}\msun$ and $10^{15}\msun$ clusters.  The separate radiation
  components are labeled in the figures. Emission from the primary electrons
  are represented by the dashed curves, and emission from secondary electrons
  and positrons and $\pi^{0}$ $\gamma$-rays are represented by the dotted
  curves.  The solid curves are the total nonthermal photon spectra.  Panels
  (a) and (b) show the nonthermal spectrum at $0.8\gyr$ ($z=0.22$), and panel
  (c) at $3.42\gyr$ ($z=0.0$) after the merger shock forms.  Magnetic field
  strengths are $B=1.0\mug$ for panels (a) and (c), and $B=0.1\mug$ for panel
  (b).  Panels (a) and (b) correspond to times before the termination of the
  nonthermal particle injection.  Panel (c) is approximately $2.3\gyr$ after
  particle injection has ceased.
 \label{fig:total_spectrum}}
\end{center}
\end{figure}

Separate radiation components produced by nonthermal particles are shown in
Figure \ref{fig:total_spectrum} for a shock formed in the merger of
$10^{14}\msun$ and $10^{15}\msun$ clusters, with the interaction beginning at
$z_i =0.3$. The system has been evolved for 0.8 Gyr ($z = 0.22$) in panels (a)
and (b), and for 3.42 Gyr ($z = 0$) in panel (c). Panels (a) and (b)
correspond to times before particle injection has stopped.  The cluster
magnetic field strength $B = 1.0\mug$ in panels (a) and (c), and $B = 0.1\mug$
in panel (b).  The radiation components of the primary and secondary electrons
are shown by the short-dashed and dotted curves, respectively, and the $\pi^0$
decay $\gamma$-rays are shown by the dotted curves. The curves are labeled by
the various radiation processes, and the total spectral energy distributions
are shown by the solid curves. The synchrotron and Compton radiations from
primary electrons dominate the emission at times before primary injection
stops, though $\pi^0$ decay $\gamma$ rays make a dominant contribution above
100 TeV. At late times after particle injection ceases, the cooling primary
electrons provide the dominant contribution at the lowest radio frequencies
and at optical/UV frequencies, but secondary electrons can make the dominant
contribution at 1.4 GHz.  At $\gamma$-ray energies, primary bremsstrahlung and
secondary $\pi^0$-decay $\gamma$ rays make the dominant contribution at late
times. Note that structure in some of the emission components is due either to
knock-on electrons or to the separate effects of the stellar, X-ray, and CMB
Compton energy losses on the electron energy spectra.  As can be seen, the
intensity of the synchrotron radio emission is very sensitive to the magnetic
field strength.

\begin{figure}
\begin{center}
\leavevmode
\hbox{%
\epsfxsize=6.5in
\plotone{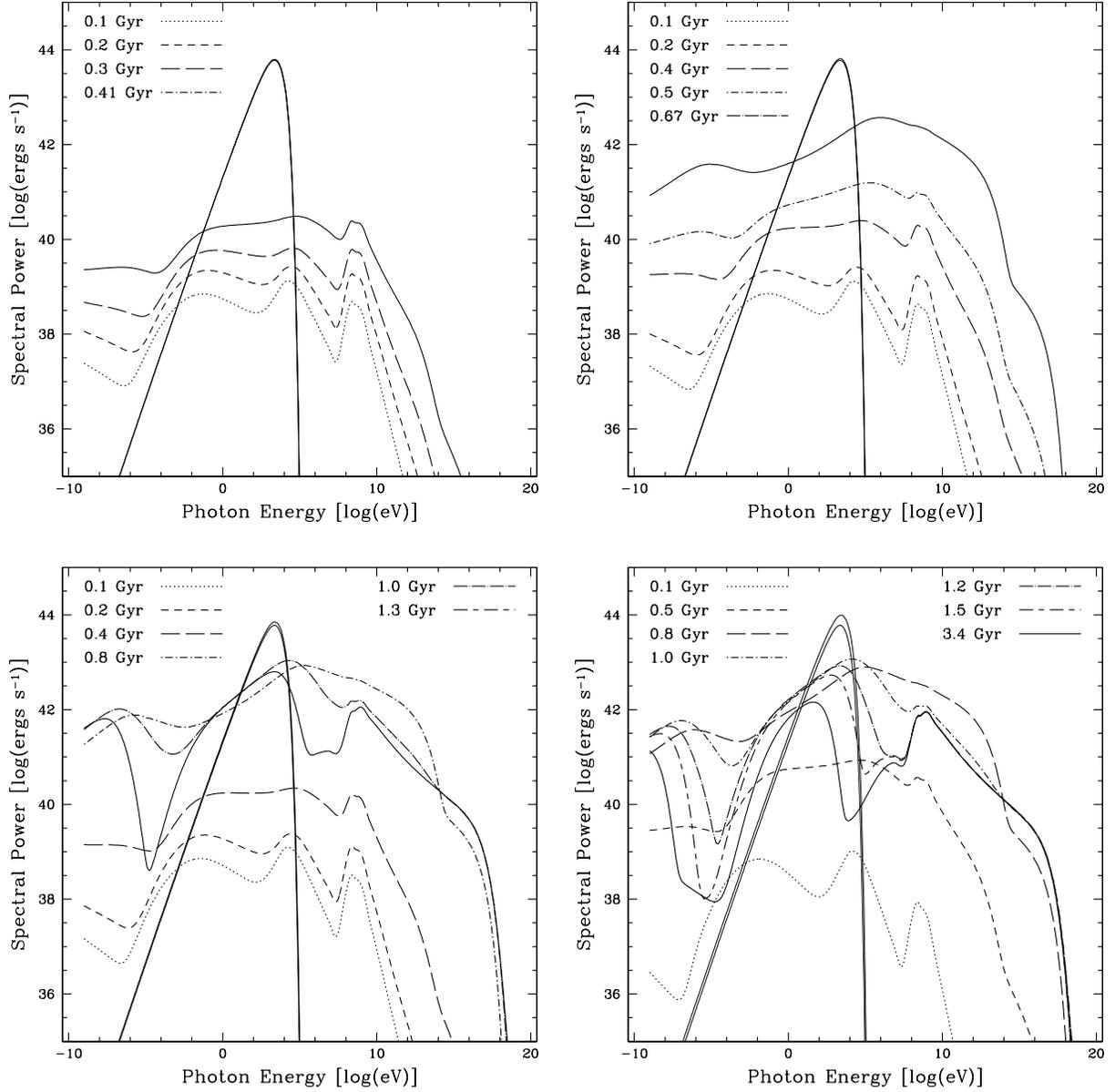}}
\caption{Temporal evolution of the total spectral energy distribution for a
  cluster merger shock which begins at redshift $z_{i}=0.03$ ({\em top left}),
  $z_{i}=0.05$ ({\em top right}), $z_{i}=0.1$ ({\em bottom left}), or
  $z_{i}=0.3$ ({\em bottom right}).  Each shock is evolved to the present
  ({\em solid line}).  Intermediate times are indicated in each panel.  In all
  cases the shock survives for $\sim 1.1\gyr$.  Cluster merger shocks that
  develop at redshifts $\gtrsim 0.1$ do not persist to the present epoch.  The
  thin solid curve shows the expected thermal bremsstrahlung for a
  $10^{15}\msun$ cluster, whose temperature is given by equation
  (\ref{eqn:cluster_temperature}).  The figures with two red curves represent
  the thermal bremsstrahlung at $z_{i}$ (lower curve) and $z=0.0$ (higher
  curve).  The integrated luminosity of the thermal bremsstrahlung is $\approx
  10^{45}\ergss$.}
\label{fig:total.time.b1.0}
\end{center}
\end{figure}

Figure \ref{fig:total.time.b1.0} shows the temporal evolution of the total
photon spectra radiated by a shock formed in the merger of $10^{14}\msun$ and
$10^{15}\msun$ clusters.  The photon spectra are shown at various times for
shocks that begin at redshifts $z_i=0.03$, $0.05$, $0.1$, and $0.3$.  Each
shock is evolved to the present time ($z=0.0$), and the radiation spectrum at
the present time is represented by the solid curve.  Typical lifetimes for the
shocks are $\approx 1\gyr$.  As a result, only shock fronts with $z \geq 0.1$
show late-time evolution of the total photon spectra.

At early times well before particle injection ceases ($\lesssim 0.5$ Gyr), the
luminosity increases as the shock front accelerates particles and the
particles accumulate.  During this epoch, the injection spectrum is very soft
and nonthermal bremsstrahlung and $\pi^0$ radiation can dominate the
production of photons at high energies because of the relatively few
high-energy electrons. The peaks in the spectra at $t<0.3\gyr$ spectra are,
from low to high energies, due to Compton, bremsstrahlung, and $\pi^0$
radiation. The enhancement at low frequencies is due to nonthermal synchrotron
radiation.  As the front ages and the electron spectrum hardens ($0.5$-$1$
Gyr), the number of high-energy electrons increases and the Compton process
strongly dominates the nonthermal emission at infrared energies and above.

At times between $t_{\rm coll}$ and $t_{\rm acc}$ (between $\approx 0.9$ and
1.1 Gyr in Fig.\ \ref{fig:total.time.b1.0}), the shock front propagates to the
outer regions of the clusters, and the particle injection rate decreases.  The
number of high energy electrons begins to decrease as their previous numbers
are no longer replenished.  The consequence of this behavior is a softening in
the synchrotron and high-energy Compton emission.  This is seen clearly in the
$z_{i} = 0.1$ and $z_{i} =0.3$ panels.  As the Compton emission declines, the
$\gamma$ rays from the $p+p \rightarrow \pi_{0}\rightarrow 2\gamma$ process
become increasingly evident at times $\gtrsim 1.1\gyr$.

The shock front terminates at $t_{\rm acc}$ after propagating through the
cluster.  At these late times, particle injection ceases, and the remaining
population of nonthermal particles evolve according to the energy loss and
diffusion processes described in \S\S\ref{appsub:electron_loss_rates} \&
\ref{appsub:proton_loss_rates}, with no freshly accelerated electrons to
repopulate the particle distribution.  The system quickly becomes devoid of
primary electrons, so that the bremsstrahlung and pion emission dominate at
hard X-ray energies and above.  The most extreme case is shown in the $z_{i} =
0.3$ panel of Figure \ref{fig:total.time.b1.0}.  The spectrum at the present
epoch shows three maxima produced by synchrotron radiation at kHz frequencies,
Compton scattering near 100 eV, and combined nonthermal bremsstrahlung and
$\pi_{0}$-decay emission near 100 MeV. The plateau in the radio emission at
late times is formed by synchrotron emission from secondary electrons (1.4 GHz
corresponds to $6\times 10^{-6}$ eV).

Also shown in Figure \ref{fig:total.time.b1.0} is the thermal bremsstrahlung
emission from the ICM using the form for the emission spectrum given by
\citet{rybicki:79}.  The temperature and the luminosity of the thermal
bremsstrahlung is computed from equations in \S
\ref{subsub:cluster_environment}. Thermal bremsstrahlung generally dominates
the emission at $10$ eV - $10 \kev$ energies.  It should be noted that the
relations for temperature and luminosity of the thermal bremsstrahlung
component do not take into consideration the merging process of the system.
Computer simulations of expected X-ray temperatures and luminosities of
merging clusters indicate an increase in the thermal bremsstrahlung luminosity
and temperature \citep{ricker:98,schindler:93}.  Therefore the thermal
bremsstrahlung curves presented in Figs. \ref{fig:total.time.b1.0} and
\ref{fig:total.time.b0.1} very likely underestimate the true X-ray luminosity
and temperature of the gas during the merger event.  At times $\gg t_{\rm
acc}$, the relation between the initial temperature and luminosity will depend
on mass ratio of the merger components, the dark-matter density profile of the
clusters, and impact parameter of the merger event \citep{ricker:01}.

Figure \ref{fig:total.time.b0.1} shows a calculation for the same system as in
Figure \ref{fig:total.time.b1.0}, except for a cluster with a magnetic field
of 0.1 \mug.  We consider these two values because of the uncertainty
regarding the mean magnetic fields in clusters of galaxies.  Despite Faraday
rotation measurements indicating magnetic fields strengths $B\gtrsim 1.0\mug$
\citep{clarke:01}, a comparison of radio and Compton luminosities suggest
field strengths $B\approx 0.1\mug$ \citep{fusco-femiano:99}.  Possible
explanations for this discrepancy may arise from an observational biasing of
the Faraday rotation measurement of the magnetic fields--weaker magnetic
fields mean brighter Compton emission for a fixed radio luminosity--and the
possibility that the distributions of the magnetic field and nonthermal
particles are anticorrelated due to the accumulation of electrons in weak
magnetic field regions \citep{petrosian:01}. The major change between the two
figures is an increase in the synchrotron radio emission, which is roughly
proportional to the square of the magnetic field. The system with the lower
magnetic field has a slightly brighter Compton component.

Light curves at various observing frequencies are shown in Fig.\
\ref{fig:light_curves} using the parameters for our standard case given by
Fig.\ \ref{fig:total.time.b1.0} with $B = 1.0\mug$, and a cluster merger
beginning at $z_i = 0.3$. The light curves of the nonthermal radiation exhibit
a common behavior independent of frequency. At early times, the spectral power
rises rapidly as the clusters merge. The peak emission occurs when the centers
of mass of the two clusters pass at $t_{\rm coll}$, after which the emission
exhibits a slow decay and approaches a plateau at times $t \gtrsim t_{\rm
acc}$ when particle injection has stopped. The rate of decay of the emission
increases with radio frequency due to the stronger cooling of the higher
energy electrons, so that the decay is slowest at lower frequencies.
Synchrotron emission from secondary electrons forms the late-time plateaus at
radio energies.  This behavior is also apparent for the hard X-ray emission,
although it is formed by primary bremsstrahlung and both primary and secondary
Compton radiation at late times.  At $\gamma$-ray energies, the $\pi^0$-decay
emission forms a plateau of emission that dominates soon after $t = t_{\rm
acc}$.

\begin{figure}
\begin{center}
\leavevmode
\hbox{%
\epsfxsize=6.5in
\plotone{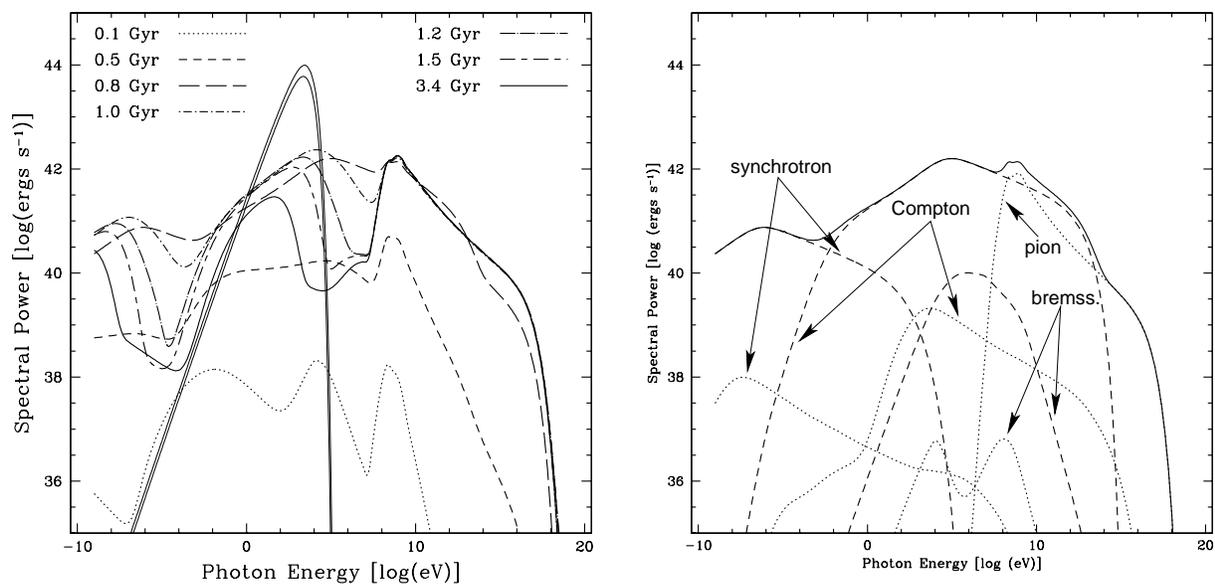}}
\caption{Same as the $z_i = 0.3$ case in Fig.\ \ref{fig:total.time.b1.0},
except that $\eta_e = 1$\% and $\eta_p = 10$\%. Right panel shows the separate
radiation components at time $t=0.8\gyr$ using the same notation as Fig.\
\ref{fig:total_spectrum}, and the left panel shows the total spectral power at
various times. }
\label{fig:efficiency}
\end{center}
\end{figure}

In Fig.\ \ref{fig:efficiency}, we show the spectral energy distributions and
spectral components for a case where the efficiency $\eta_p =10$\% for protons
and $\eta_e = 1$\% for electrons.  An enhancement of the number flux of
nonthermal protons compared to electrons is expected in nonrelativistic shock
acceleration because protons, unlike electrons, resonate with Alfv\'en waves
at all energies \citep{bar99}, and the Larmor radii of protons in the tail of
the Maxwellian distribution of the background thermal plasma is greater than
that of the electrons, so that protons can more effectively scatter across the
shock front. This theoretical expectation is borne out by the large
proton-to-electron ratio in the cosmic rays. The enhanced proton contribution
produces structure in the $\gamma$-ray spectrum even at $t\approx t_{\rm
coll}$ when the emission is brightest. The $\pi^0$-decay feature will be
clearly evident in the spectrum of merging clusters of galaxies for such a
large proton-to-electron efficiency ratio.

Calculations of the hardest particle injection spectral index $s_{\rm min}$
formed in cluster merger shocks are shown in Fig.\
\ref{fig:minimum_spectral_index} as a function of the larger mass $M_{1}$ of
the two clusters, with the subcluster mass $M_2 =10^{14}\msun$.  To derive
these results, we used equations in \S \ref{subsub:evolution_of_merger} to
calculate the dynamics of the clusters, equations in \S
\ref{subsub:cluster_environment} to calculate the sound speed, and equations
(\ref{eqn:compression_ratio}) and (\ref{eqn:injected_power_law_index}) to
calculate the spectral index from the Mach number.  We also assume that the
onset of the merger begins at redshift $z_i = 0.1$; softer injection indices
are obtained for mergers at larger values of $z_i$ because of the smaller
maximum separations $d$ at earlier times, given by equation
(\ref{eqn:turn_around_distance}). We calculate $s_{\rm min}$ for various
values of $r_c$ and $\beta$. The values $(r_c,\beta) = (0.05,0.8),
(0.05,0.45), (0.5, 0.8)$, and $(0.5,0.45)$ roughly correspond to the extrema
in the range of these parameters measured for 45 X-ray clusters observed with
{\it ROSAT} \citep{wu:00}. Also shown are values of $s_{\rm min}$ for
$(r_c,\beta) = (0.179, 0.619)$, which are the average values of these
parameters for the 45 X-ray clusters, and $(r_c,\beta) = (0.25, 0.75)$, which
are the standard parameters used in the calculations.

The uppermost curves with large $r_c$ values display a maximum.  At larger
values of $M_1$, $s_{\rm min}$ hardens due to the stronger gravitational
potential.  Although $s_{\rm min}$ spans a wide range of values, the value of
$s_{\rm min}$ corresponding to the average values of $r_c$ and $\beta$ lies in
the range $2.1\lesssim s_{\rm min}\lesssim 2.4$, and more realistically in the
range $2.2\lesssim s_{\rm min}\lesssim 2.4$ because clusters with masses
$3\times 10^{15}$-$10^{16}\msun$ are very rare. Because $z_i = 0.1$, there is
a constraint on the available time for the interaction to occur. At low values
of $M_1$, the gravitational interaction is small so that the interaction takes
a longer time to occur. The features in the curves between $10^{14}$ and
$10^{14.5}\msun$ are a result of this effect.

\section{Discussion}
\label{sec:discussion}

We have presented results of a computer simulation designed to calculate the
temporal evolution of nonthermal particles accelerated by a shock front formed
between two merging clusters of galaxies.  We then calculated the expected
total nonthermal photon spectra resulting from synchrotron, bremsstrahlung,
and Compton radiation from both primary and secondary electrons, and
$\pi_{0}$-decay $\gamma$ radiation from accelerated protons.  Our results
apply to the shocks formed in the merger of two clusters of galaxies, and not
to the accretion shocks formed by spherical infall of matter during the
process of structure formation.  Theoretical studies suggest that accretion
shocks should exist \citep{bertschinger:85,miniati:00}, and that each cluster
itself will be the center of its own accretion flow. Although X-ray and radio
observations
\citep{kassim:01,markevitch:01,fusco-femiano:02b,bagchi:02,valtchanov:02} of
halos and relics provide strong evidence for the existence of merger shocks,
there is also observational evidence for accretion shocks
\citep{ensslin:98b,ensslin:01,slee:01,ensslin:02}. We have focused on merger
shocks in this study.  Previous work by \citet{fujita:01} would classify our
simulations as accretion events, and they showed the luminosity of these
accretion shocks to be $\sim 10^{43} \ergss$ when $M_1 = 1\times10^{15}\msun$,
which is consistent with the results of our simulations.

\begin{figure}
\begin{center}
\leavevmode
\hbox{%
\epsfxsize=6.5in
\plotone{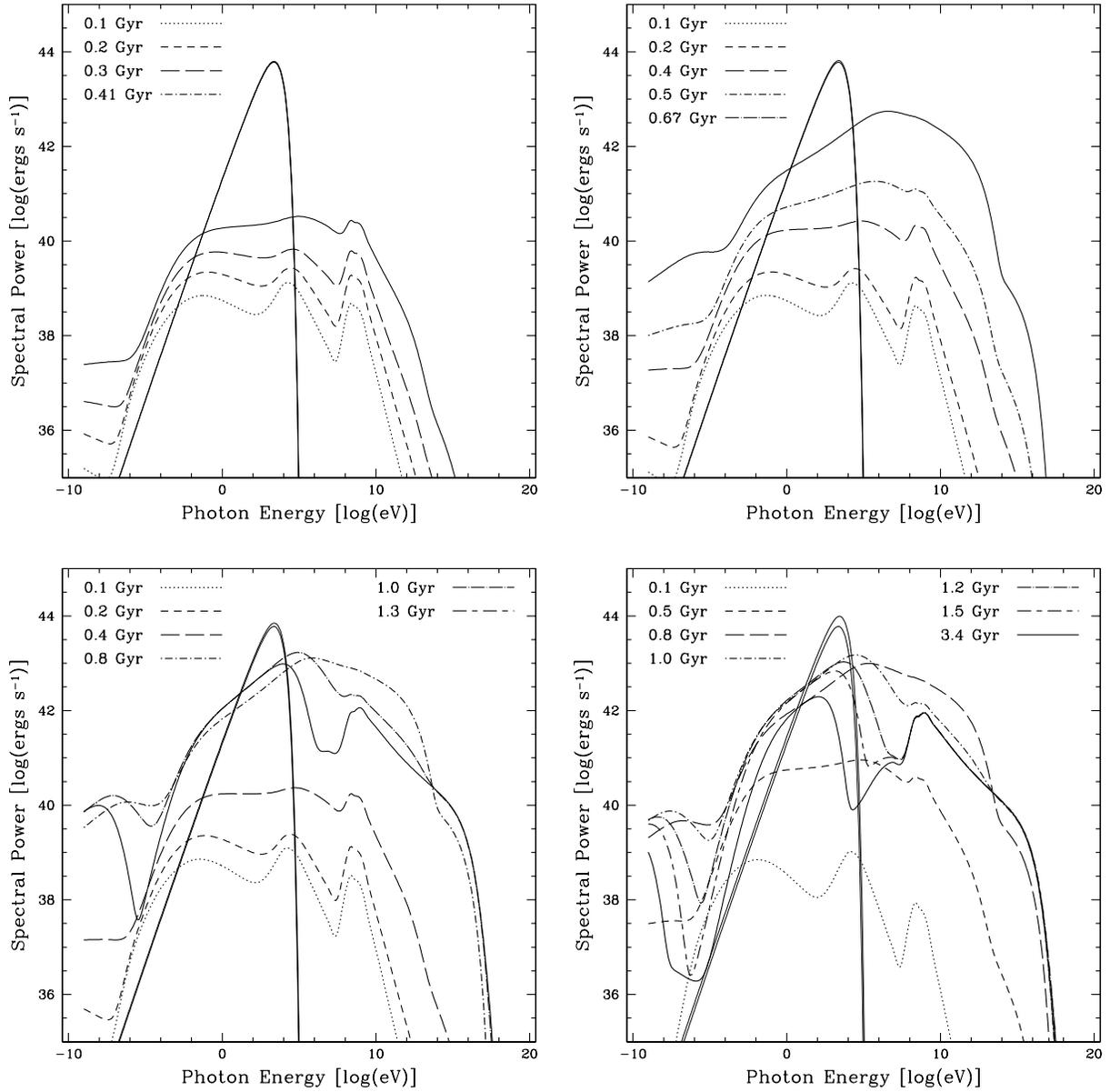}}
\caption{Same as Figure \ref{fig:total.time.b1.0}, except with an assumed
  cluster magnetic field of $0.1\mug$.}
\label{fig:total.time.b0.1}
\end{center}
\end{figure}

Beppo-SAX and Rossi X-Ray Timing Explorer (RXTE) observations show the
presence of a hard X-ray (HXR) excess at energies $\gtrsim 20\kev$ from A1656
(Coma) \citep{fusco-femiano:99,rephaeli:99}, A2256 \citep{fusco-femiano:00},
and A3667 \citep{fusco-femiano:02a}.  Our simulations support a
Compton-scattering CMB interpretation of the HXR \citep{rephaeli:79} and EUV
\citep{hwang:97,ensslin:98a,lieu:99a,av00} emission. As shown in figure
\ref{fig:total.time.b1.0} and \ref{fig:total.time.b0.1}, the nonthermal
radiation from the merger shock during the post-collision evolution ($> t_{\rm
coll}$) shows an excess above the thermal bremsstrahlung in both the HXR and
EUV regions of the electromagnetic spectrum.  For the standard parameters
used, the calculated luminosity of the nonthermal component is $L_{NT} \approx
10^{43}$ for the HXR between the $20$-$80\kev$ energies.  For the Coma
cluster, BeppoSAX observations indicate that the HXR flux is
$2.2\times10^{-11}\flux$ which corresponds to an observed luminosity of
$\sim10^{43}\ergss$ in the $20$-$80\kev$.  This observed value is consistent
with our simulations for the merger model between the times $t_{\rm coll}$ and
$t_{\rm acc}$.  Our model does match well with the calculated mass ratio of
the Coma cluster merger model \citep{colless:96}.  In A2256, the observed
masses of the merging clusters are $1.6\times10^{15}\msun$ and
$5.1\times10^{14}\msun$ \citep{berrington:02}.  The observed HXR flux is
$1.2\times10^{-11}\flux$ which corresponds to a HXR luminosity
$\sim10^{44}\ergss$\citep{fusco-femiano:00}.  By scaling our models to the
observed masses of the merging clusters, then expected HXR luminosity is $\sim
10^{44}\ergss$ which is consistent with the observed values seen in A2256.

The Extreme Ultraviolet Explorer has detected EUV emission in the
$\sim$60-250\ev\ band for a number of clusters.  These clusters include Virgo,
Coma, Fornax, A2199, A1795, and A4059
\citep{lieu:96a,lieu:96b,lieu:99a,lieu:99b,berghofer:00b,bowyer:96,
bowyer:98,mittaz:98,bonamente:01,kaastra:99}. Note, however, that no detection
of EUV emission, other than that which could be attributed to thermal tail
X-ray emission, is reported from analysis of EUVE observations for Fornax
\citep{bowyer:01}, A2199 and A1795 \citep{bb02}, and A4059
\citep{berghofer:00a}.  The measured EUV excess in the 60-250\ev\ band for
Coma and A1795 is $2\times10^{43}\ergss$\citep{lieu:96a} and
$3\times10^{43}\ergss$, respectively \citep{bonamente:01}. The calculated EUV
luminosity of the nonthermal component from our simulations is $L_{NT} \approx
10^{43}$ for these EUV energies, and is consistent with both clusters.  As
mentioned before, our simulations are a good match to the Coma cluster merger
scenario.  The masses of the merging clusters in A1795 are uncertain
\citep{oegerle:94}, and confirming the results of our simulations with the
observed luminosity of the EUV excess of A1795 is difficult.  An EUV excess of
luminosity $\sim6\times10^{42}\ergss$ was detected from the cluster A2199
\citep{kaastra:99}.  However, A2199 lacks strong evidence for a current merger
event.  If any mergers occurred in the formation history of A2199, they
occurred long enough in the past to allow the cluster to revirialize.  It is
possible that an adjacent cluster A2197 is beginning to interact with A2199
\citep{oegerle:01}, but seems unlikely that it is the origin of the EUV excess
in the central 400\kpc\ of the cluster.  Depending on the masses of the
merging clusters, the decreased luminosity of the EUV emission in A2199 is
consistent with a cooling nonthermal particle population.  Radio and X-ray
observations of the central regions of A2199 indicate it is complex
\citep{owen:98}, and it is possible other mechanisms are producing the EUV
excess emission.

We treated in detail the merger of a $10^{14}\msun$ cluster with a
$10^{15}\msun$ cluster having values of $r_c$ and $\beta$ in equation
(\ref{eqn:number_density}) approximately equal to the mean values found in a
survey of ROSAT clusters of galaxies \citep{wu:00}. We found that the hardest
spectral index reached by the merger shock in this system is $s_{\rm min}
\approx 2.2$ for $z_i = 0.1$. Because $r_c$ and $\beta$ are not correlated
with cluster temperature $T_X$ and therefore cluster mass, as can easily be
seen from the results of \citet{wu:00}, our model results are reasonably
representative of typical observed cluster mergers.

\begin{figure}
\begin{center}
\leavevmode
\hbox{%
\epsfxsize=6.5in
\plotone{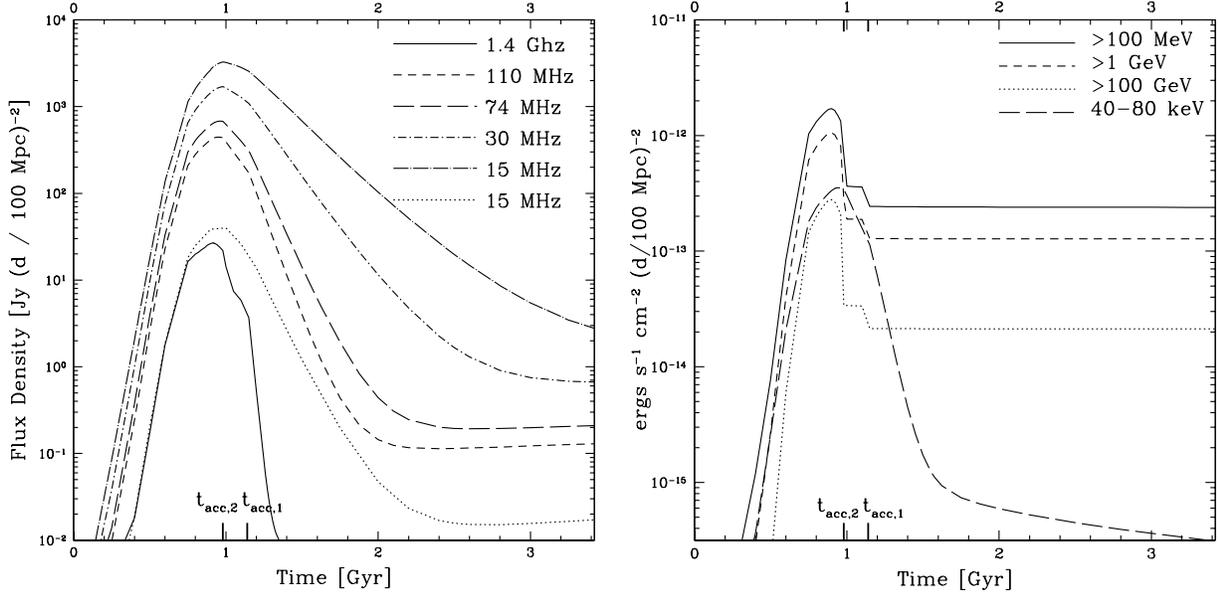}}
\caption{Light curves at various observing frequencies produced by a shock
formed in a merger between $10^{14}$ M$_\odot$ and $10^{15}\msun$ clusters
that begins at $z_i = 0.3$ and is evolved to the present epoch ($t = 3.42$
Gyr). All light curves are for a magnetic field strength of $B=1.0\mug$ unless
otherwise noted.  Radio light curves in Jansky units are given at 15 MHz, 30
MHz, 74 MHz, 110 MHz, and 1.4 GHz on the left panel, and light curves in
energy flux units are given at 40-80 keV, $> 100$ MeV, $> 1$ GeV and $> 100$
GeV in the right panel.  The 15 MHz light curve is also calculated with a
magnetic field strength of $B=0.1\mug$ ({\em dotted curve}).}
\label{fig:light_curves}
\end{center}
\end{figure}

We consider the detectability of nonthermal synchrotron emission with the
Low-Frequency Array (LOFAR\footnote{http://www.lofar.org/}). LOFAR is a
multi-element interferometric low-frequency radio telescope designed for high
resolution radio imaging in the 10-300 MHz frequency range.  Figure
\ref{fig:light_curves} includes several frequencies observable by LOFAR
showing the optimization of the LOFAR telescope for detecting these objects.
Estimated sensitivities for the LOFAR telescope range from 1.6 mJy at 15 MHz
and 40 $\mu$Jy at 110 MHz.  The quoted sensitivities are for one hour
exposures for a single beam for a total of 13,365 dual polarization dipoles
optimized in the 10-90 MHz, and 213,840 dual polarization dipoles optimized
for the 110-220 MHz each with a maximum baseline of 400 km.  With station
diameters of 65m, resolution of LOFAR is approximately $12\arcsec$ at 15 MHz,
and $2.4\arcsec$ at 110 MHz at the full extent of the array.  Because
multi-element interferometric radio telescopes have trouble detecting large,
exteneded sources, we must consider the angular size of the shock front. Given
a size scale of $\sim$1 Mpc, these features will easily be resolved at all
frequencies visible to LOFAR at a distance of 100 Mpc.  As a worst case
scenario, we will consider the extent of the array that is only capable of
observing the shock front as a point object.  For all but the most distant
shock fronts and the lowest frequencies, central 2 km of the LOFAR array,
known as the virtual core, is the optimum configuration.  Within this virtual
core is found 25\% of the total collection area of the LOFAR array.
Approximate peak luminosities taken from Figure \ref{fig:light_curves} are
$4\times10^{3}$ Jy at 15 MHz, and $5\times10^{2}$ Jy at 110 MHz.  For a shock
front with an angular size of $\approx 30\arcmin$ this gives $\sim 1$
mJy/$\sq\arcsec$ and $\sim 100\;\mu$Jy/$\sq\arcsec$, respectively.  At these
intensities, LOFAR will be able to detect any cluster merger shocks out to a
distance of $\sim 2000\mpc$ in the 15 MHz frequency band and $\sim 700\mpc$ in
the 110 MHz frequency band.  These calculations neglect any absorption due to
the interstellar medium within the Milky Way galaxy.

We use results derived using our standard system to estimate the detectability
of cluster mergers with EGRET and {\it
GLAST}.\footnote{http://glast.gsfc.nasa.gov/} Generalizing the approach of
\citet{dermer:02} to extended sources, the number of background counts
$N_{bg}$ from extragalactic diffuse $\gamma$-ray background photons with
energies $> E_\gamma$ detected within solid angle $\Delta \Omega$ of a
high-latitude source is given by
\begin{equation}
N_{bg}(>E_\gamma)\cong {(\Delta \Omega) A_0 (\Delta t)K_B E_{100}\over
\alpha_B - a_0 - 1}\; \left({E_\gamma\over E_{100}}\right)^{1+a_0-\alpha_B}\;,
\label{eqn:N_bg}
\end{equation}
where $E_{100} = 100$ MeV, $K_B = 1.72(\pm 0.08)\times 10^{-7}$ photons
cm$^{-2}$ s$^{-1}$ MeV$^{-1}$, and $\alpha_B = 2.10(\pm 0.03)$
\citep{sreekumar:98}. The on-axis effective area of the telescope is
approximated by the expression $A_0(E_\gamma/E_{100})^{a_0}$, and $\Delta t$
is the exposure-corrected time on source.

The number of source counts produced by photons with energies $> E_\gamma$ and
detected within solid angle $\Delta \Omega$ centered on the direction to a
source is given by
\begin{equation}
N_s(>E_\gamma) \cong (\Delta t)\;A_0 \int_{E_\gamma}^\infty dE \;\phi_s(E)
 \left({E\over E_{100}}\right)^{a_0} \left[1-\exp\left({-\Delta\Omega\over
 \pi\theta_d^2(E)+\pi \theta_s^2}\right)\right]\;.
\label{eqn:N_s}
\end{equation}
Here $\theta_{d}(E)$ is the energy-dependent angular resolution (point spread
function) of the telescope, $\theta_{s}$ is the angular radius of the source,
assumed to be energy-independent, and $\phi_{s}(E)$ is the source flux, which
we write as $\phi_{s}(E) = (\phi_{0}/E_{100})(E/E_{100})^{-\alpha_{s}}$
[photons cm$^{-2}$ s$^{-1}$ MeV$^{-1}$], with $E$ in units of MeV. Writing the
the $\nu L_{\nu}$ spectral power as $\nu L_{\nu} =
L_0(E^{\prime}/E^{\prime}_{100})^{\alpha_{\nu}}$, where $E^{\prime} = (1+z)E$,
$E_{100}^\prime = (1+z)E_{100}$, and $\alpha_{\nu} = 2-\alpha_{s}$ is the $\nu
L_\nu$ source spectral index, we have
\begin{equation}
\phi_s(E) = {L_0\over 4\pi d_L^2 (1.6\times 10^{-6} {\rm
ergs/MeV})\;E_{100}^2}\;\left({E\over E_{100}}\right)^{\alpha_\nu
-2}\;\cong\;{5.2\times 10^{-10} L_{43}\over d_{100}^2}\;\left({E\over
E_{100}}\right)^{\alpha_\nu -2}\;.
\label{phi_s}
\end{equation}
The luminosity distance $d_{L} = 100 d_{100}$ Mpc, and the brightest
luminosity reached at 100 MeV is $L_{0} = 10^{43} L_{43}\ergss$, noting from
the calculations in Fig.\ \ref{fig:total.time.b1.0} and
\ref{fig:total.time.b0.1} that with $L_{43} \approx 1$.

The maximum sensitivity occurs when $\Delta \Omega \cong \pi\theta_d^2(E)+\pi
\theta_s^2$, giving
\begin{equation}
N_{s}(>E_\gamma) \cong 3.3\times 10^{-8}\;{A_0({\rm cm}^2)\;\Delta t({\rm s})
L_{43}\over (1-a_0-\alpha_\nu)}\;\left({E_\gamma\over
E_{100}}\right)^{1+a_0-\alpha_B}\;.
\label{eqn:N_s1}
\end{equation}
Writing $\Delta \Omega = \pi\theta^2 = 9.6\times 10^{-4}\theta^2(^{\circ})$,
where $\theta(^{\circ})$ is the telescope's acceptance angle in degrees, the
background count rate
\begin{equation}
N_{bg}(>E_\gamma) \cong 1.65\times 10^{-8}\; {\theta^2(^\circ) A_0({\rm
cm}^2)\;\Delta t({\rm s})\over 1.1-a_0}\; \left({E_\gamma\over
E_{100}}\right)^{a_0-1.1}\;.
\label{eqn:N_bg2}
\end{equation}
The sensitivity of detection at significance $n_{\sigma}$ is therefore given
by
\begin{equation}
n_{\sigma} = {N_{s}\over \sqrt{N_{bg}} }\cong 2.6\times 10^{-4}\;{L_{43}\over
d^2_{100}}\;{ \sqrt{A_0({\rm cm}^2)\;\Delta t({\rm s})}\over \theta(^\circ)
(1-a_0-\alpha_\nu)}\left({E_\gamma\over
E_{100}}\right)^{\alpha_\nu+(a_0/2)-0.45}\;.\;
\label{eqn:nsigma}
\end{equation}
High confidence identification of a source requires that $N_s\gg 1$ and
$n_\sigma \gtrsim 5$.

For the EGRET telescope on the {\it Compton Gamma Ray Observatory}, $A_0 \cong
1200$ cm$^2$, $a_0 \cong 0$, and a nominal observing time is $\Delta t = 10^6
t_6$ s, with $t_6 \cong 1$. A source with a radius of 0.5 Mpc will have an
angular radius $\theta_s \sim 0^\circ \!.3/d_{100}$. The point spread function
of EGRET is $\theta_d(E) \cong 5^\circ (E/E_{100})^{1/2}$, which is larger
than the source angular size except for the nearest sources or when observing
at the highest energies, in which case the number of sources photons is small.
Substituting into equations (\ref{eqn:N_bg2}) and (\ref{eqn:nsigma}) gives
$N_s(>E_\gamma) \simeq 40 t_6 L_{43}(E_\gamma/E_{100})^{\alpha_\nu
-1}/[(1-\alpha_\nu)d^2_{100}]$ and $n_\sigma \simeq 1.9 \sqrt{t_6}
L_{43}(E_\gamma/E_{100})^{\alpha_\nu +0.05}/[(1-\alpha_\nu)d_{100}^2]$. Unless
$L_{43} \gg 1$, we conclude that it is unlikely that EGRET could have detected
nonthermal emission from merging clusters of galaxies.

\begin{figure}
\begin{center}
\leavevmode
\hbox{%
\epsfxsize=6.5in
\epsffile{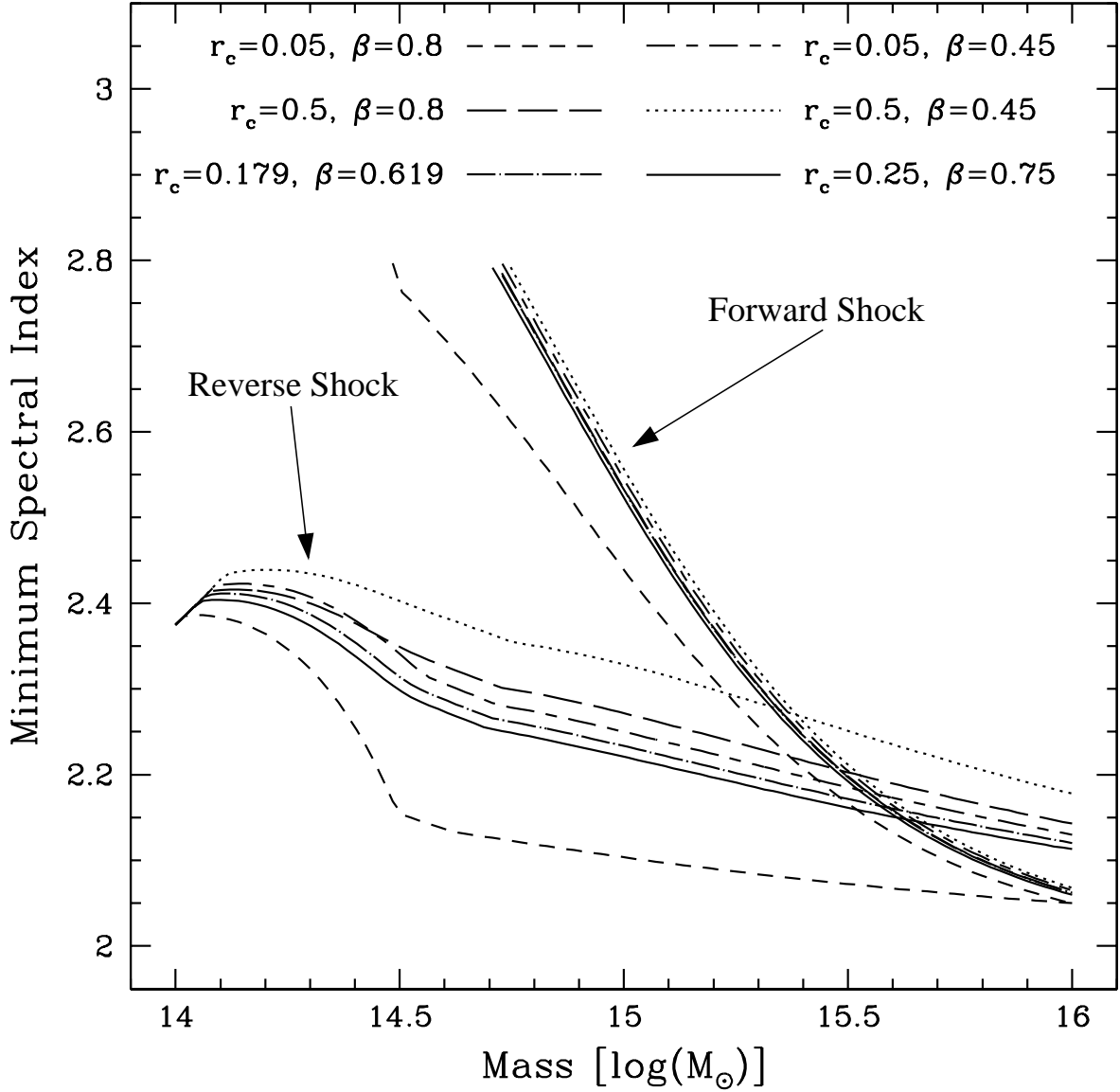}}
\caption{Calculations of the hardest particle injection spectral indices
$s_{\rm min}$ formed in cluster merger shocks as a function of the larger mass
$M_1$ of the two clusters, for various values of $r_c$ and $\beta$.  The
values of $r_c$ are given in $\mpc$.  The minimum spectral index for the
forward and reverse shock is shown in the figure. }
\label{fig:minimum_spectral_index}
\end{center}
\end{figure}

This is in disagreement with the estimate of \citet{totani:00}, who argue that
a large fraction of the isotropic unidentified EGRET sources could be due to
the radiation from merging clusters of galaxies. \citet{totani:00} make very
optimistic assumptions about the hardness of the spectral index of injected
nonthermal electrons, and in fact claim that the merger shocks have such large
Mach numbers that the injection index is nearly equal to 2.0. As shown in
Fig.\ \ref{fig:minimum_spectral_index}, such hard spectral indices are only
possible in the most centrally-peaked clusters, but that the average cluster
merger shock as inferred from observations of galaxy clusters has a minimum
injection index between 2.2 and 2.3.  Because the energies of electrons which
Compton scatter the CMB radiation to $>100$ MeV energies exceed $\sim 200$
GeV, the steeper injection spectrum reduces the available power in these
electrons by 1-2 orders of magnitude, thus accounting for the difference
between our expectations and the claims of \citet{totani:00},
\citet{kawasaki:02}, and \citet{colafrancesco:01}. Although it is possible
that a few of the unidentified EGRET sources are associated with merging
clusters of galaxies, these would involve the less frequent events involving
collisions of clusters with masses near $10^{15}\msun$, and those with hard
spectral indices or dark matter density profiles with strong central peaks.

We have not taken into account effects of a merger tree on the evolution of
the nonthermal particle distribution \citep{gabici:03}. The protons that
radiate in the EGRET and {\it GLAST} ranges will accumulate, and could
contribute factors of 2-4 enhancements due to successive mergers, but not
enough to significantly alter our conclusions. Our predictions are in accord
with a recent analysis of EGRET data \citep{rei03} (see, however,
\citet{sm02}) that fails to detect gamma-ray emission from galaxy clusters
that emit the most luminous thermal X-ray emission.

For the Large Area Detector on {\it GLAST}, $A_0 \cong 6200$ cm$^2$, $a_0
\cong 0.16$, and the on-source observing time for a scanning mode lasting
$t_{yr}$ years is $\Delta t \cong 0.2 \times 3.16\times 10^7 t_{yr}$ s, where
the factor 0.2 is the exposure correction. The point spread function for {\it
GLAST} is $\theta_{d} \cong 3^{\circ}\!.5(E/E_{100})^{-2/3}$, which still
generally dominates the source angular size except at the highest energies and
for the closest sources.  From this we obtain $N_s(>E_\gamma) \simeq 1300
\;t_{yr} L_{43}(E_\gamma/E_{100})^{\alpha_\nu -0.84} /
[(0.84-\alpha_\nu)d_{100}^2]$ and $n_{\sigma} \simeq 14\; L_{43}
\sqrt{t_{yr}}(E_{\gamma}/E_{100})^{\alpha_\nu +0.3} /
[(0.84-\alpha_{\nu})d_{100}^{2}]$.  These estimates indicate that clusters of
galaxies within $\approx 200$ Mpc will be detected with {\it GLAST}, but we
predict far fewer clusters of galaxies than predicted by \citet{totani:00} for
the reasons discussed above. Note that the detection significance is only
weakly energy-dependent, so that the criterion for detection becomes the
requirement of detecting a few photons which is more easily satisfied at lower
energies.

Our results also have relevance to suggestions that clusters of galaxies
contribute to the diffuse $\gamma$-ray background. \citet{loeb:00} have argued
that Compton-scattered CMB radiation from relativistic electrons accelerated
by shocks formed during structure formation can make a significant
contribution to the diffuse extragalactic $\gamma$-ray background, which has a
featureless power-law spectrum between a few MeV to $\sim 100$ GeV
\citep{sreekumar:98}. If this emission is provided by cluster merger shocks,
then our results indicate that this contribution cannot be significant for two
reasons: first, the density distributions inferred from X-ray observations of
thermal bremsstrahlung in Abell clusters imply an average minimum injection
spectral index that would produce Compton-scattered CMB emission that is
softer than the spectrum of the diffuse $\gamma$-ray background. Although
those few clusters with the hardest indices would contribute the most emission
above 100 MeV, the superposition of the emission from the softer sources would
produce a concave spectrum at lower energies, which is not observed. Secondly,
an associated hadronic signature would be observed even in the case that
protons are accelerated with the same efficiency as the electrons at all times
except near $t_{coll}$ (see Figs.\ \ref{fig:total.time.b1.0} and
\ref{fig:total.time.b0.1}).  The hadronic feature would, however, be seen in
the time-averaged $\gamma$-ray spectrum, so that calculations omitting
hadronic emissions are incomplete \citep{kes03}.  If, as argued in \S
\ref{sec:results}, the hadronic acceleration efficiency exceeds that of
electrons, then a $\pi^{0}$-decay signature should be apparent in the
background radiation (see Fig.\ \ref{fig:efficiency}) unless accretion shocks
make the dominant source of high-energy $\gamma$ ray cluster emission.  In
this case, most of the $\gamma$-ray emission should be located on the
periphery of the cluster, and the $\pi_{0}$ $\gamma$-rays would be reduced
dramatically due to the decreased lower gas density in the proximity of the
protons (see \citet{min03} for a spatially resolved model including both
proton and electron emissions).

Should it be the case that cluster merger shocks {\it do} produce a large
fraction of the diffuse extragalactic $\gamma$-ray background, then both low
hadronic efficiencies and dark matter halos with strong central density peaks
in the majority of rich clusters would seem to be required (see Fig.\
\ref{fig:minimum_spectral_index}). Under the assumptions of isothermality
assumed by the violent relaxation of dark matter halos which collapse from
uniform expansion of the Hubble flow, dark matter halos should not have
central density cusps. Furthermore, X-ray temperature and luminosity profiles
are inconsistent with strongly peaked central dark-matter density profiles
\citep{makino:98,wua:00} despite the predictions of N-body simulations
\citep{navarro:97}.  Clusters of galaxies could still make a small
contribution of order $\sim\!\!1$-10\% to the extragalactic diffuse
$\gamma$-ray background radiation \citep{colafrancesco:98,min02}.

Our simulations show that protons may reach $\sim 10^{19}$ eV for standard
parameters, but are not accelerated to higher energies, in agreement with the
gyroradius estimates of \citet{nma95} for maximum cosmic ray energies.  The
maximum energies of protons are primarily constrained by the available
acceleration times at early times. At later times, the proton Larmor radii may
exceed the size scale of the shock, so that the protons then diffuse into the
surrounding galaxy cluster of galaxies and undergo no further acceleration.
Competition of the acceleration and $p\gamma$ energy-loss rates plays a small
role in limiting the highest energy of protons for our standard parameters.

For a diffusion coefficient corresponding to a Kolmogorov spectrum of
turbulence, $\gtrsim 10^{16}\ev$ protons can diffuse from the cluster on a
Hubble time \citep{berezinsky:97}. Assuming Bohm diffusion, however,
\citet{berezinsky:97} and \citet{volk:96} showed that clusters of galaxies
provide a storage room for cosmic rays with energies $\lesssim 10^{18}$ eV on
cosmological time scales.  Our calculations confirm that $\sim\!\!10^{18}\ev$
protons remain within the cluster for times much greater than the lifetime of
the shock front (see Fig.\ \ref{fig:time_energy_spectra}).  For energies
$\gtrsim 10^{19}\ev$, the loss of protons from diffusion through the ICM will
become significant over the lifetime of the shock front, and will become a
restriction to the maximum energies.  The maximum energy imposed by the Larmor
radius size-scale criterion is always less than the maximum energy determined
by the diffusion loss timescale for our models.  For a given magnetic field,
the Larmor radius of nonthermal particles is weakly dependent on the cluster
mass ($\propto M^{1/3}$), and despite the possible range of masses for
$M_{1}$, the maximum particle energy $E_{\rm max} \propto M^{1/3}$, and varies
little over the mass range of $M_{1}$. \cite{kang:97} argue that more rapid
particle acceleration can take place in perpendicular shocks, but this
requires preferential orientations of the magnetic field with respect to the
shock direction.

\section{Summary and Conclusions}

We have presented the results of a computer code designed to calculate the
nonthermal particle distributions of a shock front formed in the merger event
of two clusters of galaxies.  We have calculated nonthermal particle energy
spectra for primary electrons and protons, and secondary electrons.  Photon
spectra were calculated for bremsstrahlung, Compton, and synchrotron processes
as well as $\pi^{0}$-decay $\gamma$ radiation from pp collisions.  Our results
apply to shocks that form at the interaction boundary between two merging
clusters of galaxies, and not to cluster accretion shocks which form at the
outer regions of the cluster.  Evidence for nonthermal particle acceleration
in merger shocks, which penetrate into the inner region of the accreting
cluster, are provided by radio halos and relics and nonthermal X-ray emission
from clusters of galaxies.

The thermal X-ray bremsstrahlung was modeled from observations of
luminosities, temperatures, and masses of clusters of galaxies. We modeled the
nonthermal emission under the assumption that electrons and protons are
accelerated with a 5\% efficiency of the available gravitational energy that
is dissipated during the course of gravitational interactions between two
merging clusters of galaxies.  Particle acceleration at shocks formed between
merging clusters of galaxies produces a population of nonthermal electrons
that Compton-scatters CMB photons. This process naturally produces a power-law
distribution of photons in the 40-80\kev\ energy range with luminosities
consistent with HXR observations of the galaxy clusters A1656, A2256, A3667
and, possibly, A2199.  In addition, the EUV emission observed in A1656 is also
consistent with a nonthermal Compton-scattered CMB origin.  Cluster magnetic
field strengths $\sim\!\!0.1\mug$ are in accord with this interpretation of
the observed HXR excess and EUV emission.

In contrast to the results of \citet{totani:00}, \citet{kawasaki:02}, and
\citet{colafrancesco:01}, we have argued that it is unlikely that more than a
few of the isotropic unidentified EGRET sources can be attributed to radiation
from nonthermal particles produced by cluster merger shocks.  Previous studies
have assumed hard nonthermal spectra that give brighter $\gamma$-ray emission
than implied by our simulation results.  Such hard spectra can only be
obtained in the infrequent merger events involving two very massive clusters,
or between clusters where the dark-matter density profiles are centrally
peaked.

The diffuse extragalactic $\gamma$-ray background is a featureless power law
with photon index of $2.10(\pm 0.03)$.  The dominant nonthermal $\gamma$
radiation components in cluster-merger shocks include Compton-scattered CMB
radiation and bremsstrahlung from nonthermal electrons, and $\pi^{0}$-decay
emission from nuclear interactions involving nonthermal protons. The $\pi^0$
decay signature will be present if the hadronic acceleration efficiency
exceeds the electron acceleration efficiency, as expected in diffusive shock
acceleration theory. Furthermore, spectra calculated using parameters obtained
from ROSAT observations of 45 Abell clusters \citep{wu:00} produce spectra
with slopes $\approx 2.2$-$2.4$.  Unless dark matter density profiles are
centrally peaked and hadronic acceleration efficiency is low, our results
imply that nonthermal emission from merging clusters of galaxies can make only
a minor contribution to the diffuse extragalactic $\gamma$-ray
background. Using standard diffusive shock acceleration with mean ICM magnetic
fields $\lesssim 1 \mu$G, our results also indicate that the merger shocks in
clusters of galaxies do not accelerate $\gtrsim 10^{19}$ eV cosmic rays,
though additional effects, such as shock obliquity and the presence of
preexisting particle populations, could permit higher energy acceleration by
these shocks.

\acknowledgements {We wish to thank Armen Atoyan, Namir Kassim, and Paul Ray
for useful discussions, and the referee for a detailed and constructive
report. We would also like to thank Steve Sturner for providing his supernova
code which was used to begin this project. This work is supported by the
Office of Naval Research and NASA Astrophysical Theory Grant DPR S-13756-G and
NASA GLAST Science Investigation Grant DPR S-15634-Y.}

\appendix

\section{Particle Energy Loss Rates}
\label{app:particle_energy_loss_rates}

\subsection{Electron Loss Rates}
\label{appsub:electron_loss_rates}

Nonthermal electrons lose energy through Coulomb or ionization interactions
with background thermal electrons.  Energy is lost through this process at the
rate
\begin{equation}
\dot{K}_{e,{\rm coul}}(E_{e},t) = - \eta^{e}_{\rm He}~ n_{\rm ICM} \left(
\frac{4 \pi e^{4} \Lambda(t)} {\beta_{e} m_{e} c} \right) \left(\Xi(t) -
\frac{d\Xi(t)}{dx}\right),
\label{eqn:coulomb_energy_loss}
\end{equation}
where we have made use of the following definitions:
\begin{equation}
\Xi(t) = 2 \pi^{-\frac{1}{2}} \int^{x(t)}_{0} dy ~ y^{\frac{1}{2}} \exp(-y),
\end{equation}
\begin{equation}
\Lambda(t) = 24 - \ln \left[ \frac{(\eta^{e}_{\rm He}~n_{\rm
ICM})^{\frac{1}{2}}}{T^{\rm eV}_{e} (t)} \right]\mbox{, and}
\label{eqn:coulomb_logarithm}
\end{equation}
\begin{equation}
x(t) = \frac{m_{e} v^{2}_{e}}{2kT_{e}(t)}.
\end{equation}
The quantity $n_{ICM}$ is the proton density of the ICM, $\beta_{e}c$ is the
electron velocity, $T_{e}(t)$ is the electron temperature given in Kelvin, and
k is Boltzmann constant.  Nonthermal electrons diffuse in energy space
according to the diffusion coefficient for Coulomb scattering
\begin{equation}
D_{e,{\rm coul}}(K_{e}, t) = \eta^{e}_{\rm He}~ n_{\rm ICM} \left( \frac{8 \pi
e^{4} \Lambda(t)} {\beta_{e} m_{e} c} \right) k T_{e}(t) \Xi(t)
\label{eqn:coulomb_diffusion_coefficient}
\end{equation}
\citep{miller:96}.

For a fully ionized ICM, the bremsstrahlung energy loss rate
\citep{blumenthal:70} is given by the expression
\begin{equation}
\dot{K}_{e,{\rm brem}}(K_{e},t) = - \eta^{n}_{\rm He}~ n_{\rm ICM} \left(
\frac{8 e^{6}}{m^{2}_{e} c^{4} \hbar} \right) (\ln \gamma_{e} + 0.36)(K_{e} +
m_{e} c^{2}) \;,
\label{eqn:bremsstrahling_energy_loss}
\end{equation}
where $\eta^{n}_{\rm He} = 1.3$ is an enhancement factor for Helium that
results from the $Z(Z+1)$ dependence of the loss rate.  The synchrotron energy
loss rate for electrons averaged over pitch angle in a magnetic field of
strength $B$, Lorentz factor $\gamma_{e}$, and velocity $\beta_{e}$ is
\citep{rybicki:79}
\begin{equation}
\dot{K}_{e,{\rm syn}}(E_{e},t) = -\frac{4}{3} c \sigma_{T} \gamma^{2}_{e}
\beta^{2}_{e} \left( \frac{B^{2}}{8 \pi} \right),
\label{eqn:synchrotron_energy_loss}
\end{equation}
where $\sigma_{T}$ is the Thomson cross section.  

Electrons lose energy by Compton scattering ambient photons in the ICM
environment, as described by equations (\ref{eqn:xray_energy_density}),
(\ref{eqn:stellar_energy_density}), and (\ref{eqn:CMB_energy_density}) for the
cluster X-ray, stellar, and CMB radiation fields, respectively.  The spectral
distribution of each radiation field is approximated by a blackbody spectrum
of temperature $T_{X}$ (see equation (\ref{eqn:cluster_temperature})), $4000$
K, and $2.7 (1 + z)$ K at epoch $z$.  The Compton energy loss rate is given by
\begin{equation} 
\dot{K}_{e,{\rm comp}}(E_{e}, t) = - \frac{4}{3} c \sigma_{T} \gamma^{2}_{e}
\beta^{2}_{e} \sum_{i} U_{i} \sigma_{\rm KN}(\gamma_{e} \vartheta_{i})\;
\label{eqn:compton_energy_loss}
\end{equation}
\citep{skibo:93}.  The quantity $\sigma_{KN}(\gamma_{e}, \vartheta_{i}) \equiv
\sigma_{C}(\gamma_{e} \vartheta_{i}) / [\sigma_{T}(1 + \gamma_{e}
\vartheta_{i})]$ is a Klein-Nishina correction factor, where $\vartheta_{i}
\equiv kT_{i} / m_{e} c^{2}$ is the dimensionless temperature, and
$\sigma_{C}(\epsilon_{\gamma})$ is the Compton scattering cross section for
photons with energy $\epsilon_{\gamma} = h\nu / m_{e} c^{2}$ in the electron
rest frame.

Electrons can also lose energy through the double (or ``radiative'') Compton
process in which a second photon is created during an electron-photon
interaction. Estimates by \citet{rw71} and \cite{gou75} show that the ratio of
energy-loss rates through double Compton scattering and ordinary Compton
scattering is $\sim\!\! 1/4$ when the typical photon energy in the electron
rest frame is$\sim 50$ MeV. This ratio only reaches unity when the rest frame
photon energy is $\sim 5$ TeV. Synchrotron energy losses completely dominate
at these energies, so the energy-loss rate through double Compton scattering
can be safely neglected. A detailed treatment of electron energy losses in
galaxy clusters is given by \citet{sar99}.

\subsection{Proton Loss Rates}
\label{appsub:proton_loss_rates}

The Coulomb energy loss rate for protons is given by
\begin{equation}
\dot{K}_{p,{\rm coul}}(K_{p},t) = - \eta^{e}~ n_{\rm ICM}
\left(\frac{4 \pi e^{4} \Lambda(t)}{\beta_{p} m_{p} c} \right)
\left[\left(\frac{m_{p}}{m_{e}}\right) \Xi(t) - \frac{d\Xi(t)}{dx}\right].
\label{eqn:proton_coulomb_energy}
\end{equation}
In addition to energy losses, the protons diffuse in energy space according to
the diffusion coefficient 
\begin{equation}
D_{p,{\rm coul}}(K_{p},t) = \eta^{e}~ n_{\rm ICM} \left(\frac{8 \pi
e^{4} \Lambda(t)}{\beta_{p} m_{p} c} \right) kT_{e}(t) \Xi(t)
\label{eqn:proton_diffusion_coefficient}
\end{equation}
\citep{huba:94}.

Protons undergo proton-proton collisions to produce pions in which a proton
loses $\approx \onethird$ of its energy per collision.  After three
collisions, the proton is assumed to be lost from the system. The timescale
for this loss is
\begin{equation}
\tau_{\rm pion}(E_{p},t) = \frac{1}{c \beta_{p} n_{\rm ICM} (\sigma_{\rm H} +
0.1 \sigma_{\rm He})}
\label{eqn:pion_catastrophic_loss}
\end{equation}
\citep{sturner:97}, where $\sigma_{\rm H}$ and $\sigma_{\rm He}$ are the
inelastic cross sections for pion-producing collisions with protons and
helium, respectively.  Measured values for these cross sections are
$\sigma_{H} = 28\mb$ and $\sigma_{He} = 100\mb$ well above threshold,
respectively \citep{meyer:72}.

High-energy protons interact with the CMB photons due to photomeson
production.  It requires about 5 interactions for protons to lose a
substantial fraction of their energy due to interactions with photons with
proton rest-frame energies in the range $200\mev$-$500\mev$, and about 2
interactions for photons whose energies in the rest frame of the protons are
greater than $500\mev$.  The time scale for the catastrophic loss is given by
\begin{equation}
\tau^{-1}_{p\gamma \rightarrow \pi}(E_{p},t) = \frac{c}{2} \int_{0}^{\infty}
d\epsilon~ \frac{n_{\gamma}(\epsilon)}{(1+z)^{3}} \int_{-1}^{1} d\mu~ (1 -
\mu)~ k_{p\gamma}(\epsilon^{\prime})\sigma_{p\gamma \rightarrow
\pi}(\epsilon^{\prime})\;
\label{eqn:pgamma_catastrophic_loss}
\end{equation}
where $\sigma_{p\gamma \rightarrow \pi}(\epsilon^{\prime})$ is the cross
section of the proton-photon interaction, and $\epsilon^\prime = \gamma_{p}
\epsilon (1 - \mu)$ is the photon energy in the rest frame of the proton. Here
$\mu = \cos \theta$, and $\theta$ is the angle between the directions of the
interacting photon and proton.  We let $\sigma_{p\gamma \rightarrow
\pi}(\epsilon^{\prime}) = 380\;\mu$barns and 120 $\mu$barns, and the
inelasticity coefficient $k_{p\gamma}=0.2$ and 0.6, for $200\mev \leq
\epsilon^\prime \leq 500\mev$, and $\epsilon^\prime > 500\mev$, respectively
\citep{atoyan:01}.  The number density of CMB photons at redshift $z$ is
denoted by $n_{\gamma}(\epsilon)$.

Protons diffuse through the ICM at a rate that depends on the the diffusion
coefficient.  This diffusion process is a random walk which allows particles
to escape from the system.  Protons that escape from the cluster merger are
treated as a catastrophic loss with time scale \citep{volk:96}
\begin{equation}
\tau_{d}(E_{p}) \approx R_{1}^{2} / \kappa_{\rm Bohm}(E_{p}),
\label{eqn:diffusion_catastrophic_loss}
\end{equation}
where we employ the B\"{o}hm diffusion coefficient $\kappa_{\rm Bohm}(E_{p})=
\beta c r_{\rm L}/3$ for particle diffusion, and $r_{\rm L}$ is the particle
Larmor radius.

\subsection{Pion Production from pp-Collisions}
\label{appsub:secondary_electron_production}

Nonthermal protons collide with ambient gas particles to produce secondaries
through several channels.  The five dominant channels are (1) p + p
$\rightarrow \pi^0 + X$;(2) p + p $\rightarrow \pi^+ + X$;(3) p + p
$\rightarrow \pi^- + X$;(4) p + p $\rightarrow K^+ + X $;(5) p + p
$\rightarrow K^- + X$.  Here ``$X$'' refers to any other byproducts produced
in the reaction other than the particle indicated. It is also understood that
channel (2) does not include the contribution resulting from the production of
deuterium in the reaction p + p $\rightarrow \pi^{+} + d$, which can be
treated separately but is a small contribution to the total secondary
production.  The secondary electron production is dominated by the first two
pion channels, and are the only two channels considered in our calculations.
Our calculation technique follows the method described by
\citet{dermer:86a,dermer:86b} and \citet{moskalenko:98}; see also
\citet{bc99}.  This model combines the isobaric model \citep{stecker:70} at
energies $<\!\!3\gev$ and the scaling model \citep{badhwar:77} for energies
$>\!\!7\gev$.  We use a linear combination to join the two models in the
transition region between 3\gev\ and 7\gev.

For proton energies $<\!\!3\gev$, the production of pions produced by
proton-proton collisions is mediated by the excitation of a $\Delta_{3/2}$
isobar.  Assuming the outgoing $\Delta_{3/2}$-isobar of mass $m_{\Delta}$ is
collinear with the initial direction of the colliding protons in the
center-of-momentum system (CM) with isotropically distributed decay products
in the laboratory system (LS), the pion distribution in the LS of the isobaric
model is given by
\begin{equation}
f(K_{\pi}; K_{p}, m_{\Delta}) = \frac{1}{ 4 m_{\pi} \gamma_{\pi}^{\prime}
\beta_{\pi}^{\prime} \gamma_{\Delta}^{+} \beta_{\Delta}^{+} } \left\{
H[\gamma_{\pi}; \aleph^{+}(-), \aleph^{+}(+)] +
\frac{\gamma_{\Delta}^{+}\beta_{\Delta}^{+}}
{\gamma_{\Delta}^{-}\beta_{\Delta}^{-}} H[\gamma_{\pi}; \aleph^{-}(-),
\aleph^{-}(+)] \right\}
\end{equation}
where the Heavyside function $H[x;a,b]$ is defined to be $=1$ if $a\leq x \leq
b$; otherwise $=0$.  The function $\aleph^{\pm}(\mp) = \gamma^{\pm}_{\Delta}
\gamma^{\prime}_{\pi} (1 \mp \beta^{\pm}_{\Delta} \beta^{\prime}_{\pi})$, and
$T_{\pi}$ is the pion LS kinetic energy.  The forward (+) and backward (-)
moving isobars Lorentz factors are $\gamma^{\pm}_{\Delta} = \gamma_{c}
\gamma^{\star}_{\Delta}(1 \pm \beta_{c} \beta_{\Delta}^{\star})$, where
$\gamma_{c} = \sqrt{s} / 2 m_{p}$ is the Lorentz factor of the CM with respect
to the LS, and $\gamma^{\star}_{\Delta} = (s + m_{\Delta}^{2} + m_{\pi}^{2}) /
2 s^{1/2} m_{\Delta}$ is the Lorentz factor of the isobar in the CM.  The pion
Lorentz factor in the rest frame of the $\Delta$ isobar is
$\gamma^{\prime}_{\pi} = (m^{2}_{\Delta} + m^{2}_{\pi} - m^{2}_{p})/2
m_{\Delta} m_{\pi}$.

The pion spectrum resulting from the pp collisions is calculated by evaluating
the following integral over the isobar mass spectrum:
\begin{equation}
\frac{dN(K_{\pi},K_{p})}{dT_{\pi}} = w_{r}(K_{p}) \int^{\sqrt{s} -
m_{p}}_{m_{p} + m_{\pi}} dm_{\Delta} ~ B_{\cal W}(m_{\Delta}) f(K_{\pi}; K_{p},
m_{\Delta}).  
\end{equation}
$B_{\cal W}(m_{\Delta})$ is the normalized isobar mass spectrum given by the
Breit-Wigner distribution
\begin{equation} 
B_{\cal W}(m_{\Delta}) = \frac{w_{r}(K_{p}) \Gamma} {(m_{\Delta} -
\langle m_{\Delta} \rangle)^{2} + \Gamma^{2}}
\end{equation}
with average isobar mass $\langle m_{\Delta} \rangle$, and normalization
factor
\begin{equation}
w_{r}(K_{p}) = \left[ \arctan\left( \frac{\sqrt{s} - m_{p} - \langle
m_{\Delta} \rangle}{\Gamma}\right) - \arctan \left( \frac{m_{p} + m_{\pi} -
\langle m_{\Delta} \rangle }{\Gamma} \right) \right]^{-1}.
\end{equation}
Our calculations only consider the $\Delta_{3/2}(1232)$ isobar which has a
resonance width $\Gamma = 1/2 \times 115\mev$ \citep{particle:84}.

For protons whose energies are greater than $7\gev$, we adopt the scaling
model of \citet{stephens:81}.  The Lorentz invariant cross section given by
\citet{badhwar:77} and \citet{stephens:81} for neutral and charged pion
production is given by
\begin{equation}
E_{\pi} \frac{d^{3} \sigma}{d^{3} p_{\pi}} = {\cal A} {\cal
G}_{\pi}(E_{p})(1-\zeta_{\pi})^{\cal Q} \exp[-{\cal B} p_{\perp} /
(1+4m_{p}^{2}/s)],
\end{equation}
where we have made use of the following definitions:
\begin{eqnarray}
{\cal G}_{\pi^{\pm}}(E_{p}) = (1 + 4m_{p}^{2}/s)^{\cal R},\\
{\cal G}_{\pi^{0}}(E_{p}) = (1 + 23 E_{p}^{-2.6})(1 - 4m_{p}^{2}/s)^{\cal R},\\
{\cal Q} = (C_{1} - C_{2} p_{\perp} + C_{3} p_{\perp}^{2}) / \sqrt{1 +
4m_{p}^{2}/s},\\ 
\zeta_{\pi} = \sqrt{x^{\star}_{\parallel} + (4/s)(p_{\perp}^{2} +
m^{2}_{\pi})},\\
x^{\star}_{\parallel} = \frac{2 m_{\pi} \gamma_{\pi} \gamma_{c}
s^{\frac{1}{2}} (\beta_{\pi} \cos \theta - \beta_{c})}{\sqrt{(s - m_{\pi}^{2}
- m_{X}^{2})^{2} - 4 m_{\pi}^{2} m_{X}^{2}}}.
\end{eqnarray}
The constants ${\cal A}$, ${\cal B}$, $C_{1,2,3}$, and ${\cal R}$ are given by
\citet{badhwar:77} and \citet{stephens:81}, and $\theta$ is the pion LS polar
angle.  Adopted values for the constants ${\cal A}$, ${\cal B}$, $C_{1,2,3}$,
and ${\cal R}$ are given in Table \ref{tab:invariant_cross_section_constants}.
The quantity $m_{X}$ depends on the channel under consideration and are (1)
$m_{X} = 2 m_{p}$, (2) $m_{X} = m_{p} + m_{n}$, (3) $m_{X} = 2m_{p} +
m_{\pi}$, (4) $m_{X} = m_{p} + m_{n}$, and (5) $m_{X} = 2 m_{p} + m_{K}$.

To calculate the LS energy distribution of pions, we integrate the following
over the pion LS polar angle:
\begin{equation}
Q_{\pi}(E_{\pi}, E_{p}) = \frac{2 \pi p_{\pi}}{\langle \eta
\sigma_{\pi}(E_{p}) \rangle} \int^{1}_{\cos \theta_{\rm max}} d \cos
\theta ~ \left( E_{\pi} \frac{d^{3} \sigma}{d^{3} p_{\pi}} \right).
\label{eqn:pion_distribution}
\end{equation}
where $-1 \leq \cos \theta_{\rm max} \leq 1$ and is defined as 
\begin{equation}
\cos \theta_{\rm max} = \frac{1}{\beta_{c} \gamma_{c} p_{\pi}} \left(
\gamma_{c} E_{\pi} - \frac{s - m_{X}^{2} m_{\pi}^{2}}{2 \sqrt{s}} \right).
\end{equation}
The quantity $\langle \eta \sigma_{\pi}(E_{p}) \rangle$ is the inclusive cross
section and its values are taken from equation (5-9) of \citet{dermer:86a}.

The pions will decay to produce muons.  These muon are created fully
polarized.  This results in a $e^{\pm}$ decay asymmetry.  The muons created by
the pion decay are created in the rest frame of the pion with Lorentz factor
$\gamma_{\mu} = (m^{2}_{\pi} + m^{2}_{\mu}) / 2 m_{\pi} m_{\mu} \approx
1.039$, and $\beta_{\mu} \approx 0.2714$.  In the rest frame of the muons, the
resulting electron decay spectrum is given by \citet{commins:73,orth:76}
\begin{equation} 
\frac{d^{2}N}{d E^{\star}_{e}~ d \cos \theta^{\star}} = \frac{2\epsilon^{2}(e -
2\epsilon)}{m_{\mu}} \left[ 1 + \xi \left( \frac{1-2\epsilon}{3-2\epsilon}
\right) \cos \theta^{\star} \right],
\end{equation}
where $E_{e}^{\star}$ is the electron energy in the muon rest frame and helps
to define the quantity $\epsilon = 2 E_{e}^{\star} / m_{\mu}$.  The angle
between the polarization angle of the parent muon and the resulting electron
momentum is $\theta^{\star}$.  The quantity $\xi = \pm 1$ for $\mu^{\pm}
\rightarrow e^{\pm}$.

\begin{figure}
\begin{center}
\leavevmode
\hbox{%
\epsfxsize=6.5in
\plotone{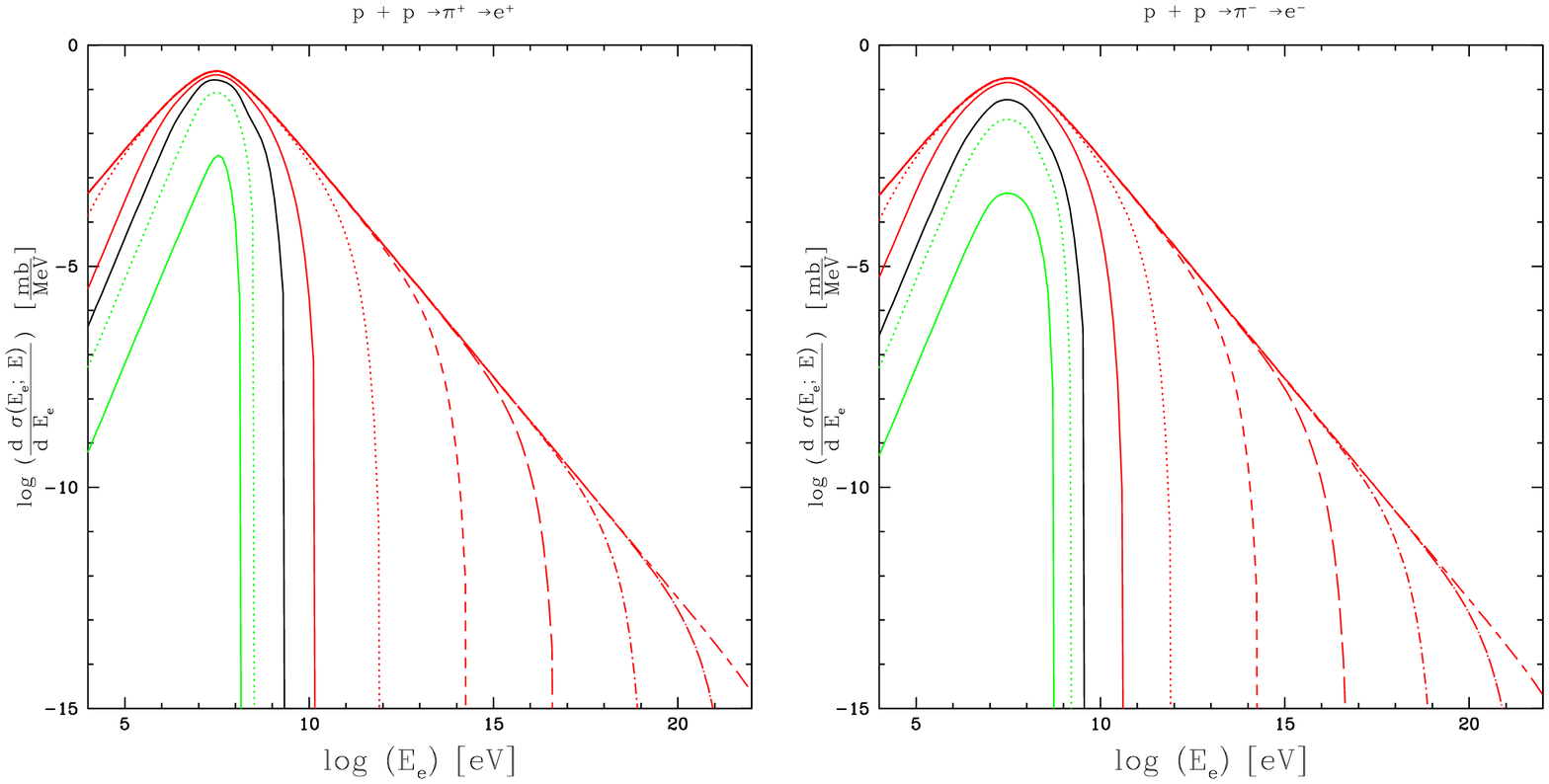}}
\caption{Energy distributions of secondary positrons (\em left panel) and
electrons (\em right panel) produced in p-p nuclear collisions by protons with
energies $K_{p} (\rm eV) =\{ 3.16\times 10^{8},5.62\times 10^{8},3.16\times
10^{9},2.15\times 10^{10},1.0\times 10^{12},2.5\times 10^{14},6.3\times
10^{16},1.58\times 10^{19},4.0\times 10^{21},1.0\times 10^{24} \}$ {\em (from
bottom to top)} for the $p+p \rightarrow e^{+} + X$ channel and protons of
energies $K_{p}=\{ 1.0\times 10^{9},2.15\times 10^{9},5.0\times
10^{9},5.0\times 10^{10},1.0\times 10^{12},2.5\times 10^{14},6.3\times
10^{16},1.58\times 10^{19},4.0\times 10^{21},1.0\times 10^{24}\ev \}$ {\em
(from bottom to top)} for the $p+p \rightarrow e^{-} + X$ channel.
\label{fig:secondary_electron_cross_section}}
\end{center}
\end{figure}

For a given electron energy $E_{e}$, the normalized electron decay energy
distribution in the frame where the pions are isotropic is \citep{dermer:86a}
\begin{equation}
\frac{dN(E_{e}; y_{\pi})}{dE_{e}} ~ =~ \left\{
\begin{array}{ll}
\phi(\varsigma_{-}) - \phi(\varsigma_{+}), & \varsigma_{+} < \varsigma_{-} <
\omega_{-},\\
\psi(\varsigma_{-}) - \psi(\omega_{-}) + \phi(\omega_{-}) -
\phi(\varsigma_{+}), & \varsigma_{+} < \omega_{-} < \varsigma_{-} <
\omega_{+},\\
\psi(\omega_{+}) - \psi(\omega_{-}) + \phi(\omega_{-}) - \phi(\varsigma_{+}),
& \varsigma_{+} < \omega_{-} < \omega_{+} < \varsigma_{-},\\
\psi(\varsigma_{-}) - \psi(\varsigma_{+}), & \omega_{-} < \varsigma_{+} <
\varsigma_{-} < \omega_{+},\\
\psi(\omega_{+}) - \psi(\varsigma_{+}), & \omega_{-} < \varsigma_{+} <
\omega_{+} < \varsigma_{-},\\
0, & \omega_{+} < \varsigma_{-},\\
\end{array}
\right.
\label{eqn:secondary_electron_function}
\end{equation}
where we have introduced the following definitions:
\begin{equation}
\begin{array}{rcl}
\varsigma_{\pm} &=& \frac{E_{e}}{\gamma_{\mu} \gamma_{\pi} m_{\mu} (1 \pm
\beta_{\pi})}\\
\omega_{\pm} &=& (1 \pm \beta_{\mu}) /2\\
\phi(x) &=& \frac{8 \gamma^{5}_{\mu} x^{2}}{dE^{\star}_{e}~ d\cos
\theta^{\star}} \left[ \frac{(3- u \beta^{2}_{\mu}) (1-\beta^{2}_{\mu})}{2} -
\frac{4x(3+\beta^{2}_{\mu} - 4 u \beta^{2}_{\mu})}{9}\right],\\
\psi(x) &=& \frac{1}{6\beta_{\mu} \gamma_{\mu} \beta_{\pi} \gamma_{\pi}
m_{\mu}} \left[ (5+u)\ln x - \frac{6(u+2u\beta_{\mu} + 3)
x^{2}}{(1+\beta_{\mu})^{2}} + \frac{16 (u + 3u\beta_{\mu} + 2)
x^{3}}{3(1+\beta_{\mu})^{3}} \right],
\end{array}
\end{equation}
where $u = \xi / \beta_{\mu}$, and $\xi = 1$ for positrons, and $\xi = -1$ for
electrons.  Figure \ref{fig:secondary_electron_cross_section} shows the energy
distributions of electrons and positrons formed through pp collisions by
mono-energetic protons with a variety of energies for channels (1) and (2)
(compare \citet{murphy:87}).

The resulting electron or positron spectrum formed by the decay of
isotropically produced pions is
\begin{equation}
Q_{e}(E_{e}) = \int^{\infty}_{\bar{\gamma}_{\pi}(E_{e})} d\gamma_{\pi}~
Q_{\pi}(\gamma_{\pi}) ~\frac{dN(E_{e}; y_{\pi})}{dE_{e}}
\end{equation}
with $Q_{\pi}(\gamma_{\pi})$ given by equation (\ref{eqn:pion_distribution}),
and $\bar{\gamma}_{\pi} = (E_{e}/E^{\rm max}_{e} + E^{\rm max}_{e}/E_{e})/2$
if $E_{e} > E^{\rm max}_{e} = m_{\mu} (1+\beta_{\mu})\gamma_{\mu}/2 \approx
69.8\mev$; otherwise, $\bar{\gamma}_{\pi} = 1$.  

\subsection{Knock-On Electrons}
\label{appsub:knock-on_electrons}

Nonthermal protons interact with thermal electrons through Coulomb collisions
to produce a population of nonthermal knock-on electrons.  The knock-on
electron source function in electrons cm$^{-3}$ sec$^{-1}$ is given by

\begin{equation}
Q_{k}(\gamma_{e})\cong 1.75~n_{ICM}(r)~4 \pi \int^{\infty}_{\gamma_{1}} d
\gamma_{p} ~ \Phi_{p\rightarrow e}(\gamma_{e}, \gamma_{p}) ~\frac{dJ_{p}}{d
\gamma_{p}}
\label{eqn:knock_on_electron_emissivity}
\end{equation}
\citep{abraham:66} where the factor 1.75 accounts for the presence of heavier
nuclei in the ICM and cosmic rays, $\gamma_{e}$ is the Lorentz factor of the
electron, $\gamma_{p}$ is the Lorentz factor of the proton, and $dJ_{p}/d
\gamma_{p}$ is the differential cosmic-ray proton intensity in particles
cm$^{-2}$ sec$^{-1}$ sr$^{-1}$.  The differential probability for the
production of an electron having a total energy per unit rest-mass energy in
$d\gamma_{e}$ at $\gamma_{e}$ by the collision of a cosmic-ray proton of
Lorentz factor $\gamma_{p}$ in units of cm$^2$ s$^{-1}$ is given by
\begin{equation}
\Phi_{p\rightarrow e}(\gamma_{e}, \gamma_{p})~d \gamma_{e} = \frac{2\pi
r_{e}^2}{(1-\gamma_{p}^{-2})} \left[ \frac{1}{(\gamma_{e} - 1)^{2}} -
\frac{m_{e} \left( \gamma_{p} + \frac{m_{e}^{2} + m_{p}^{2}}{2 m_{p} m_{e}}
\right)}{m_{p} (\gamma_{e} - 1) \gamma_{p}^{2}} + \frac{m_{e}^{2}}{2 m_{p}^{2}
\gamma_{p}^{2}}\right]~ d \gamma_{e}
\label{eqn:knock_on_electrons_cross_section}
\end{equation}
\citep{abraham:66} where $r_{e}$ is the classical radius of the electron.  The
maximum transferable energy is given by
\begin{equation}
\gamma_{\rm max} = 1 + \frac{\gamma_{p}^{2} - 1}{\frac{m_{e}}{m_{p}} \left(
\gamma_{p} + \frac{m_{e}^{2} + m_{p}^{2}}{2m_{e}m_{p}} \right) }
\end{equation}
The lower bound of the integration in equation
(\ref{eqn:knock_on_electron_emissivity}) is determined by solving the
inequality $\gamma_{e} \leq \gamma_{\rm max}$ for $\gamma_{p}$.  The result is
\begin{equation} 
\gamma_{p} \geq \gamma_{1} = \frac{m_{p}}{2 m_{e}} (\gamma_{e} - 1) + \left[1
    + \frac{1}{2}\left(1+ \frac{m_{p}^2}{m_{e}^2}\right)(\gamma_{e} - 1) +
    \frac{m_{p}^{2}}{4 m_{e}^{2}} (\gamma_{e} - 1)^{2}\right]^{1/2}.
\label{eqn:gamma_lower_bound}
\end{equation}

\section{Photon Production}
\label{app:photon_production}

Nonthermal photons are produced by electron synchrotron radiation,
electron-electron and electron-nucleon bremsstrahlung, and Compton-scattered
CMB, stellar, and X-ray radiation fields.  These components are calculated for
both primary electrons and secondary electrons and positrons.  Neutral
pion-decay $\gamma$ rays are also produced in proton-nucleon interactions. We
discuss each component in the order mentioned above.

\subsection{Synchrotron Radiation}
\label{appsub:synchrotron_radiation}

We define the synchrotron radiation emissivity as $Q_{\rm synch} = dN_{\rm
synch} / dV~dt~dE_{\gamma}$. It is calculated from the expression
\begin{equation}
\label{eqn:synchrotron}
Q_{\rm synch}(E_{\gamma}, t) = \left( \frac{4 \pi \sqrt{3} e^{3} B}{h^{2} \nu
m_{e} c^{3}} \right) \int_{0}^{\pi/2} d\theta~ \sin^{2} (\theta)
\int_{K_{e,{\rm min}}}^{K_{e,{\rm max}}} dK_{e}~ J_{e}(K_{e}, t)~
F(\nu/\nu_{c}),
\end{equation}
where $\theta$ is the electron pitch angle, $E_{\gamma} = h\nu$ is the photon
energy of frequency $\nu$, $\nu_{c} = 0.42 \gamma_{e}^{2} B_{-7} \sin
(\theta)$ in units of Hz, and $J_{e}(E_{e}, t)$ is the electron intensity 
(units of electrons cm$^{-2}$ s$^{-1}$ sr$^{-1}$ $E_e^{-1}$).  The function
$F(\nu/\nu_{c})$ is defined through the expression
\begin{equation}
\label{eqn:bessel_synchrotron}
F(\nu/\nu_{c}) = \left(\frac{\nu}{\nu_{c}}\right) \int_{\nu/\nu_{c}}^{\infty}
dx~ K_{5/3} (x),
\end{equation}
where $K_{5/3}$ is the modified Bessel function of order 5/3.  The integration
is performed numerically with approximate expressions for $F(\nu/\nu_{c})$
given by \citet{ginzburg:65}. Electrons are assumed to be isotropically
distributed in pitch angle.  In addition, we include the effects of free-free
absorption of the synchrotron spectra in the source using the absorption
coefficient \citep{ginzburg:65}
\begin{equation}
\label{eqn:ff_absorption}
\alpha_{\rm ff} ({\rm cm}^{-1}) = 10^{-2} \frac{n_{\rm ICM}^2}
{\nu^{2} T_{e}^{3/2}}\left[ 17.7 +
\ln\left(\frac{T^{3/2}_{e}}{\nu}\right) \right].
\end{equation}
where the electron temperature $T_{e}$ is given in Kelvin, the ICM particle
number density is in units of $\percc$, and the frequency $\nu$ is in Hertz.

\subsection{Bremsstrahlung Radiation}
\label{appsub:bremsstrahlung_radiation}

We calculate the electron-electron ($e$-$e$) and electron-nucleon ($e$-$n$)
bremsstrahlung emissivity by the expression
\begin{equation}
\label{eqn:bremsstrahlung_radiation}
Q_{{\rm brem},e-e,e-n}(E_{\gamma},t) = 4 \pi n_{\rm ICM} \Delta^{e,n}_{\rm He}
\int_{K_{e,{\rm min}}}^{K_{e,{\rm max}}} dK_{e}~ J_{e}(K_{e}, t)
\left(\frac{d\sigma}{dE_{\gamma}}\right)_{e-e,p},
\end{equation}
where we have included the correction factor $\Delta^{e,n}_{\rm He}$ for the
presence of Helium.  Approximate values for the Helium correction factor are
$\Delta^{e}_{\rm He} = 1.2$ and $\Delta^{n}_{\rm He} = 1.4$.  The expressions
for the differential cross sections are adopted from equation (A1) in
\citet{huag:75} for electron-electron interactions and from equation (3BN) in
\citet{koch:59} for electron-nucleon interactions.

\subsection{Compton Scattering}
\label{appsub:compton_scattering}

We calculate the Compton scattering emissivity using the full Klein-Nishina
cross section \citep{jones:68,blumenthal:70} for relativistic electrons
$(\gamma \gg 1)$ by the expression 
\begin{eqnarray}
Q_{{\rm comp},j}(E_{\gamma},t) & = & 8 \pi r_{0}^{2}(m_{e} c^{2})^{2}
\int_{0}^{\infty} dE_{t}~ \frac{n_{j}(E_{t})}{E_{t}} \int_{E_{e}^{{\rm
thresh}}(E_{t})}^{\infty} dE_{e} ~ E^{-2} J_{e}(E_{e},t) \nonumber \\* & &
\left[ 2q \ln (q) + (1 + 2q)(1-q) + \frac{(q \epsilon_{e,t})^{2}(1-q)}{2(1+q
\epsilon_{e,t})}
\right].
\end{eqnarray}
We define
\begin{equation}
\epsilon_{e,t} = \frac{4 E_{e}E_{t}}{(m_{e} c^{2})^{2}},
\end{equation}
\begin{equation}
q = \frac{E_{\gamma}}{\epsilon_{e,t}(E_{e} - E_{\gamma})},
\end{equation}
\begin{equation}
E_{e}^{{\rm thresh}}(E_{t}) = \frac{E_{\gamma} }{2} \left( 1 + \sqrt{1 +
\frac{(m_{e} c^{2})^{2}}{E_{\gamma} E_{t}}} \right){\rm , ~~and}
\end{equation}
\begin{equation}
n_{j}(E_{t}) = \frac{15 U_{j} E_{t}^{2}}{(\pi k T_{j})^{4}} \left[\exp
\left(\frac{E_{t}}{kT_{j}}\right) - 1\right]^{-1}.
\end{equation}
The energy $E_{t}$ is target photon energy, and $E_{e}^{{\rm thresh}} (E_{t})$
is the minimum energy required by an electron to scatter a photon from $E_{t}$
to energy $E_{\gamma}$.  The temperature, number density, and energy density
of the $j^{th}$ photon component is given by $T_{j}$, $n_{j}$, and $U_{j}$,
respectively, from \S \ref{subsub:cluster_environment}.

\subsection{Pion-Decay $\gamma$ Rays}
\label{appsub:pion_decay_gamma_rays}

We use the method developed by \citet{dermer:86b} to calculate the
$\pi^{0}$ emissivity from proton-proton collisions.  The $\pi^{0}$ emissivity
is calculated by 
\begin{equation}
Q^{p-p}_{\pi^{0}} (K_{\pi^{0}}, t) = 4 \pi n_{\rm ICM}
\int_{K_{p}^{{\rm thresh}}}^{\infty} dK_{p} ~ J_{p}(K_{p}, t)~ \frac{d\sigma
(K_{\pi^{0}}, K_{p})}{dK_{\pi^{0}}}
\end{equation}
where we have defined the neutral pion kinetic energy as $K_{\pi^{0}}$, and
the differential cross section for $\pi^{0}$ production is calculated from the
cross section given by \citet{dermer:86b} which uses the isobar model of
\citet{stecker:73} for proton energies less than 3\gev\ and the scaling model
of \citet{stephens:81} for protons of energies greater than 7\gev.  A linear
combination of the two models is assumed between 3 and 7 GeV.  The
$\gamma$-ray emissivity from $\pi_{0}$ decay is calculated from the expression
\begin{equation}
Q^{p-p}_{\pi^{0},\gamma}(E_{\gamma}, t) = 2 \int^{\infty}_{E_{\gamma} +
m_{\pi^{0}}^{2} c^{4}/4 E_{\gamma}} dE_{\pi^{0}} ~
\frac{Q^{p-p}_{\pi^{0}}(E_{\pi^{0}},t)}{\sqrt{E^{2}_{\pi^{0}} -
m^{2}_{\pi^{0}}c^{4}}}
\end{equation}
Considering collisions with heavy nuclei, \citet{dermer:86b} has shown that
the $\gamma$-ray emissivity is enhanced by the multiplicative factor $1.45$.\@
The total $\gamma$-ray emissivity therefore becomes
\begin{equation}
Q^{\rm total}_{\pi^{0},\gamma} = 1.45 Q^{p-p}_{\pi^{0},\gamma}
\end{equation}


\clearpage

\clearpage

\clearpage

\clearpage

\clearpage

\clearpage

\clearpage

\clearpage

\clearpage

\clearpage

\begin{deluxetable}{cccc}
\tablecolumns{4}
\tablewidth{0pt}
\tablecaption{Adopted values for constants defining the invariant cross
  section for $p+p \rightarrow X$ reactions
  \label{tab:invariant_cross_section_constants}}
\tablehead{
  \colhead{Constant} & 
  \multicolumn{3}{c}{Channel}\\ 
  \colhead{} & 
  \colhead{(1)} &
  \colhead{(2)} &
\colhead{(3)}
}
\startdata
${\cal A}$  & 140  & 153    & 127 \\
${\cal B}$  & 5.43 & 5.55   & 5.3 \\
$C_{\rm 1}$ & 6.1  & 5.3667 & 7.0334 \\
$C_{\rm 2}$ & -3.3 & -3.5   & -4.5 \\ 
$C_{\rm 3}$ & 0.6  & 0.8334 & 1.667 \\
${\cal R}$  & 2  & 1      & 3  
\enddata
\end{deluxetable}

\end{document}